\documentclass[12pt]{article}
\usepackage{float} 
\usepackage{hyperref}
\usepackage{tabularx}
\usepackage{makecell}
\usepackage{booktabs}
\usepackage{caption}
\usepackage{newfloat} 
\usepackage{placeins} 
\usepackage{siunitx}

\usepackage{tabularx,booktabs,makecell}
\newcolumntype{Y}{>{\raggedright\arraybackslash}X}

\usepackage{newtxtext,newtxmath}

\usepackage{graphicx}

\usepackage[letterpaper,margin=1in]{geometry}

\renewenvironment{abstract}
	{\quotation}
	{\endquotation}

\date{}

\makeatletter
\renewcommand{\fnum@figure}{\textbf{Figure \thefigure}}
\renewcommand{\fnum@table}{\textbf{Table \thetable}}
\makeatother

\usepackage{scicite}

\usepackage{url}

\def\scititle{Heterogeneous Ensemble Enables a Universal Uncertainty Metric for Atomistic Foundation Models}
\title{\bfseries \boldmath \scititle}

\author{%
    Kai Liu$^{1}$,
    Zixiong Wei$^{1}$,
    Wei Gao$^{2,3}$,\\
    Poulumi Dey$^{1}$,
    Marcel H.F. Sluiter$^{1}$,
    Fei Shuang$^{1\ast}$,\and
\begin{tabular}{@{}p{\textwidth}@{}}
    \raggedright 
    \small $^{1}$Department of Materials Science and Engineering, Faculty of Mechanical Engineering, Delft University of Technology, Mekelweg 2, Delft, 2628 CD, The Netherlands.\\
    \small $^{2}$J. Mike Walker’66 Department of Mechanical Engineering, Texas A\&M University, College Station, TX 77843, United States.\\
    \small $^{3}$Department of Materials Science \& Engineering, Texas A\&M University, College Station, TX 77843, United States.\\
    \small$^\ast$Corresponding authors. Emails:  F.Shuang@tudelft.nl
\end{tabular}
}

\begin{document} 

\maketitle


\newpage


\section*{Abstract}
\begin{abstract} \bfseries \boldmath
Universal machine learning interatomic potentials (uMLIPs) are reshaping atomistic simulation as foundation models, delivering near \textit{ab initio} accuracy at a fraction of the cost. Yet the lack of reliable, general uncertainty quantification limits their safe, wide‑scale use. 
Here we introduce a unified, scalable uncertainty metric \(U\) based on a heterogeneous model ensemble with reuse of pretrained uMLIPs. Across chemically and structurally diverse datasets, \(U\) shows a strong correlation with the true prediction errors and provides a robust ranking of configuration‑level risk. Leveraging this metric, we propose an uncertainty‑aware model distillation framework to produce system‑specific potentials: for W, an accuracy comparable to full‑DFT training is achieved using only \(4\%\) of the DFT labels; for MoNbTaW, no additional DFT calculations are required. Notably, by filtering numerical label noise, the distilled models can, in some cases, surpass the accuracy of the DFT reference labels.
The uncertainty‑aware approach offers a practical monitor of uMLIP reliability in deployment, and guides data selection and fine‑tuning strategies, thereby advancing the construction and safe use of foundation models and enabling cost‑efficient development of accurate, system‑specific potentials.

\end{abstract}

\newpage

\noindent
\section*{INTRODUCTION}
For decades, quantum-mechanical simulations, with density functional theory (DFT) at the forefront, have defined the benchmark for predicting materials properties. The emergence of data-driven strategies in the AI-for-Science paradigm, however, has led to machine-learned interatomic potentials (MLIPs) that achieve near-DFT accuracy with a fraction of the computational cost~\cite{behler2016perspective}. Recent advances in high-performance computing and deep-learning architectures have enabled the development of universal MLIPs (uMLIPs), or atomistic foundation models, which are trained on hundreds of millions of configurations spanning metals, organic molecules, and inorganic solids~\cite{deng2023chgnet,barroso2024open}. The field is advancing at an unprecedented pace: platforms such as Matbench Discovery now catalog more than twenty distinct uMLIP models~\cite{riebesell2025framework}, including M3GNet~\cite{chen2022universal}, CHGNet~\cite{deng2023chgnet}, MACE~\cite{batatia2023foundation}, Orb~\cite{rhodes2025orb}, SevenNet~\cite{7net}, and EquiformerV2 (eqV2)~\cite{liao2023equiformerv2}, which exhibit strong transferability across most of the periodic table and a wide range of chemical environments.

The primary application of uMLIPs lies in replacing DFT calculations for direct property prediction. However, their accuracy can degrade for specialized systems or defect-rich configurations. Systematic softening behaviors, for instance, have been reported in uMLIPs~\cite{deng2025systematic}, while predictions of surface energies, vacancy formation energies, and interface properties remain particularly challenging~\cite{focassio2024performance,CHIPS-FF}. These limitations are typically mitigated through fine-tuning on small, system-specific DFT datasets~\cite{elena2024machine,kim2025efficient}. A second challenge stems from computational efficiency: conventional uMLIPs are generally restricted to systems of thousands of atoms~\cite{CHIPS-FF}, limiting their applicability to large-scale simulations. Recent advances in model distillation have enabled the training of compact student potentials that replicate the performance of high-capacity teacher uMLIPs, preserving accuracy while accelerating inference by one to two orders of magnitude~\cite{wang2025pfd,gardner2025distillation}. Despite these promising developments, skepticism persists regarding the accuracy and reliability of uMLIPs in fully autonomous applications. This raises a critical question: how can the uncertainty of uMLIP predictions be rigorously quantified in the absence of reference DFT calculations?

Although a range of uncertainty quantification (UQ) methods exists for system-specific MLIPs (sMLIPs), which are faster than uMLIPs but typically applicable to only a small number of elements~\cite{wen2020uncertainty,peterson2017addressing,zhu2023fast,best2024uncertainty,hu2022robust,schwalbe2025model}, robust and general strategies for uMLIPs remain scarce. This represents a critical gap, as uMLIPs require reliable extrapolation across diverse chemistries and structures due to their broader deployment scope. Current probabilistic approaches show limitations: The Orb model introduces a dedicated \emph{confidence head} to estimate atomic force variances~\cite{rhodes2025orb}, while Bilbrey \textit{et al.}~\cite{bilbrey2025uncertainty} apply quantile regression within MACE to generate confidence intervals, though both methods demonstrate limited effectiveness for out-of-distribution (OOD) detection.
Feature-space distance metrics, particularly latent space distances in graph-based uMLIPs such as eqV2 and GemNet~\cite{hu2022robust,musielewicz2024improved}, show strong correlation with prediction errors. However, these methods face challenges in interpretability and scalability when applied to large, multi-element datasets. Ensemble methods have proven effective for sMLIPs~\cite{tan2023single}, but their application to uMLIPs yields mixed results. Shallow MACE ensembles can identify some OOD configurations yet systematically underestimate errors~\cite{bilbrey2025uncertainty}. The Mattersim framework~\cite{yang2024mattersim} employs five independently initialized models with identical architectures to estimate uncertainty through prediction variance, but still shows systematic underestimation. Recent work by Musielewicz \textit{et al.}~\cite{musielewicz2024improved} suggests bootstrap ensembles offer a favorable cost-accuracy balance, whereas architectural ensembles provide greater diversity at increased computational cost.

Collectively, these observations reveal the absence of a universally accepted UQ framework for uMLIPs that correlates robustly with prediction errors. The development of an uncertainty metric on an absolute, transferable scale therefore remains a pressing challenge. Addressing this challenge would bolster the safety and reliability of uMLIP deployment in critical applications while providing essential guidance for fine-tuning, model distillation, and dataset extension.

\begin{figure}[!ht]
    \centering
    \includegraphics[width=1\linewidth]{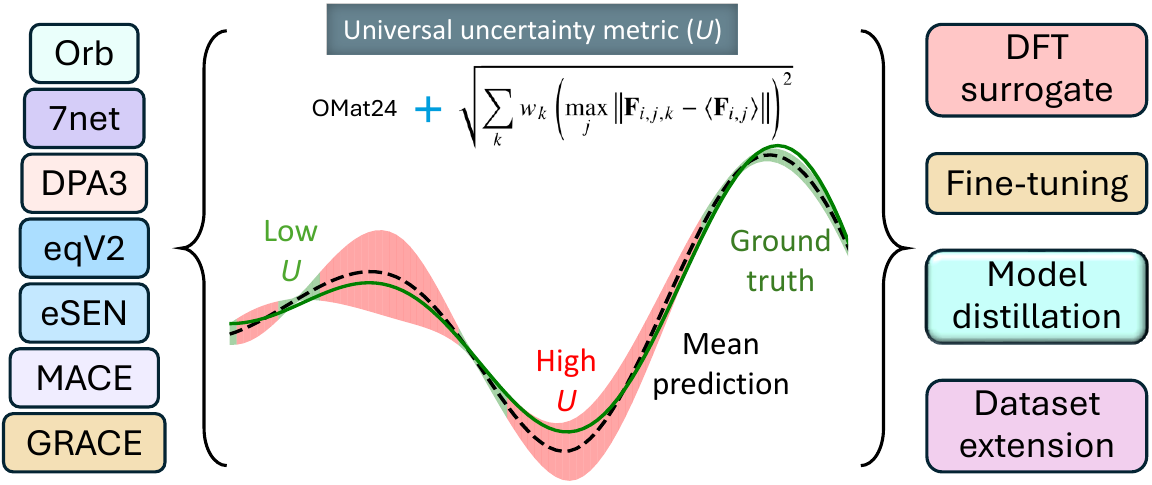}
    \caption{\textbf{The proposed universal uncertainty metric $U$ for atomistic foundation models.} A heterogeneous ensemble of models with diverse architectures constructs the uncertainty metric $U$ using the OMat24 dataset. The metric is applied to four key tasks: serving as a DFT surrogate, fine-tuning foundation models, model distillation , and extending datasets to improve foundation models.}  
    \label{Fig:1_workflow}  
\end{figure}

This work introduces a heterogeneous ensemble approach for universal UQ in uMLIPs. By strategically combining architecturally diverse uMLIPs, our method generates reliable uncertainty estimates without requiring additional training or calibration. The resulting metric exhibits strong linear correlation with prediction errors across material classes, and consistent transferability between chemical spaces. 
Comprehensive validation employs the Open Materials 2024 (OMat24) inorganic materials dataset~\cite{barroso2024open}, supplemented by systematic testing across diverse DFT-derived datasets to establish robust uncertainty thresholds. 
Practical applications demonstrate uncertainty-aware distillation of interatomic potentials for both elemental tungsten (W) and the MoNbTaW high-entropy alloy, achieving comparable accuracy to teacher models with significantly reduced computational cost. This framework provides a critical foundation for uncertainty-aware development throughout the MLIP ecosystem, enabling reliable model distillation, dataset expansion, and more trustworthy computational materials discovery, as shown in Fig.\ref{Fig:1_workflow}.

\section*{RESULTS}

\subsection*{Universal uncertainty metric $U$ via heterogeneous ensemble}

Conventional ensemble methods face fundamental scalability challenges when applied to uMLIPs. Training even one single high-accuracy uMLIP, such as eqV2 with hundreds of millions of parameters on more than 100 million atomic configurations, requires prohibitive computational resources. The challenge escalates dramatically for state-of-the-art models like Universal Models for Atoms (UMA)~\cite{UMA-meta-2025} from Meta FAIRChem, a mixture-of-experts graph network with 1.4 billion parameters trained on billions of atoms. With future uMLIPs expected to grow larger in both model size and training data, the conventional approach of training multiple independent models for UQ becomes computationally intractable. 
Conversely, academia and industry have spent millions of GPU‑hours training over twenty uMLIP architectures~\cite{riebesell2025framework}. Given the immense computational investment behind each model and the ever‑growing catalog on Matbench Discovery, developing an uncertainty metric that leverages model reuse is particularly desirable.

Here we introduce a heterogeneous ensemble framework for UQ in uMLIPs, leveraging the uMLIP models available in Matbench Discovery~\cite{riebesell2025framework}. Owing to their broad architectural and parametric diversity, the predictive accuracies of the models vary markedly (Table~\ref{tab:uMLIP}), and lower-accuracy members may introduce larger random errors that can distort ensemble estimates.
To mitigate this, we assign weights to each model proportional to its accuracy, thereby preserving ensemble diversity while limiting the influence of less reliable contributors.

This leads to a weighted formulation of uncertainty:
\begin{equation}
  U_i^{(1)}
  = \sqrt{
    \sum_{k} w_k \Bigl[\max\nolimits_{j}
      \bigl\lVert \mathbf{F}_{i,j,k}
      - \langle \mathbf{F}_{i,j}\rangle
      \bigr\rVert
    \Bigr]^2
  }\,,
  \label{Eq:weighted_UQ}
\end{equation}
where subscripts $i$, $j$, and $k$ index the configurations, atoms within a configuration, and the individual uMLIP, respectively. \(\langle \mathbf{F}_{i,j} \rangle\) denotes the average force vector. The weight $w_k$ assigned to each uMLIP model is given by
\begin{equation}
  w_k = \frac{\mathrm{RMSE}_{F,k}^{-1}}
             {\sum\nolimits_{k'=1}^{K} \mathrm{RMSE}_{F,k'}^{-1}}\,.
  \label{Eq:weights}
\end{equation}
where $\mathrm{RMSE}_{F,k}$ is the root‑mean‑square error (RMSE) in the force predictions produced by model $k$. If uniform weights \(w_k = 1/K\) are used instead, Eq.~\ref{Eq:weighted_UQ} degrades to the conventional equal‑weight uncertainty metric (denoted as $U^{(0)}$).

Additionally, we evaluate an alternative formulation that incorporates inverse‑\(\mathrm{RMSE}\) weighting during the force‑averaging step:
\begin{equation}
  U_i^{(2)}
  = \sqrt{
    \sum_{k} w_k \Bigl[\max\nolimits_{j}
      \bigl\lVert \mathbf{F}_{i,j,k}
      - \widetilde{\langle \mathbf{F}_{i,j}\rangle}
      \bigr\rVert
    \Bigr]^2
  }\,,
  \label{Eq:weighted_UQ_weighted_average}
\end{equation}
where
\begin{equation}
\widetilde{\langle \mathbf{F}_{i,j}\rangle} = \sum_k {w}_k \, \mathbf{F}_{i,j,k}.
\label{Eq:weighted_average}
\end{equation}
With these definitions, all uncertainties $U^{(0)}$, $U^{(1)}$, and $U^{(2)}$ carry units of eV/\AA, consistent with those of force and force error.

\begin{figure}[ht]
    \centering
    \includegraphics[width=1\linewidth]{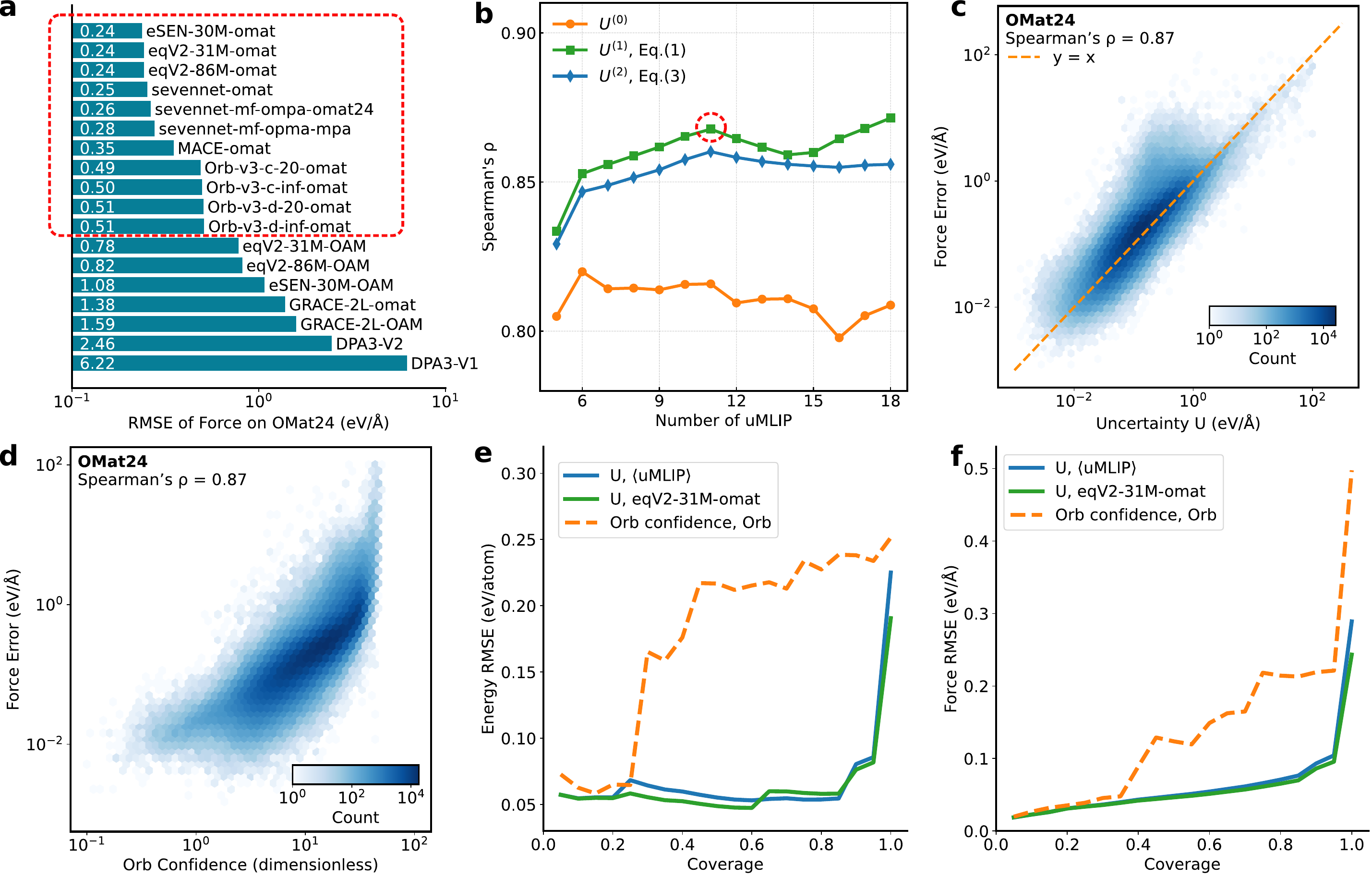}
    \caption{\textbf{Uncertainty quantification methods and their performance on the OMat24 dataset.} \textbf{a} shows names of the 18 uMLIP models used, sorted by force RMSE (low to high); more accurate models are prioritized in uncertainty estimation. \textbf{b} shows performance of three uncertainty metrics evaluated by Spearman’s $\rho$ as the number of uMLIPs varies; the selected model is marked with a red circle (as Eq.~\ref{Eq:weighted_UQ}, referred as \( U \)), and corresponding uMLIPs are highlighted in \textbf{a}. \textbf{c} is parity plot of force error vs. \( U \); color indicates point density, showing strong alignment along y = x. \textbf{d} shows force error vs. Orb-confidence (see \cite{rhodes2025orb}). \textbf{e} and \textbf{f} show force (\textbf{e}) and energy (\textbf{f}) RMSE after removing high-uncertainty configurations, as identified by \( U \) or Orb-confidence. The x-axis shows the remaining data coverage. Results are shown for both the $\langle\text{uMLIP}\rangle$ average and the efficient eqV2-31M-omat model. \( U \) leads to faster error reduction and outperforms Orb-confidence.}
    \label{Fig:2_UQ_on_OMat24}
\end{figure}

Having defined the uncertainty estimator, the next critical step is to select which uMLIP models to include in Eqs.~\ref{Eq:weighted_UQ} and \ref{Eq:weighted_UQ_weighted_average}. To ensure generality across chemistries and structures, we evaluate candidate ensembles on the public OMat24 test set, which contains more than one million configurations. Because the full OMat24 benchmark comprises over one hundred million DFT‑labeled configurations and spans a wide range of elements, bonding types, phases, and thermodynamic conditions~\cite{barroso2024open}, strong performance on its test split provides a stringent and broadly representative assessment of the generality of our uncertainty metric. We then construct the heterogeneous ensemble incrementally by ranking available models by force RMSE and adding them sequentially, beginning with the five most accurate (Fig.~\ref{Fig:2_UQ_on_OMat24}\textbf{a}). Performance is quantified using Spearman’s rank correlation coefficient \(\rho\) between predicted uncertainties and true force errors.

Fig.~\ref{Fig:2_UQ_on_OMat24}\textbf{b} shows Spearman’s \(\rho\) for \(U^{(0)}\), \(U^{(1)}\) and \(U^{(2)}\) as a function of ensemble size. For \(U^{(0)}\), optimal performance is obtained with six uMLIP models (\(\rho = 0.82\)); adding further models reduces the correlation between estimated uncertainty and true error, indicating that equal weighting allows less accurate models to degrade performance. Both \(U^{(1)}\) and \(U^{(2)}\) outperform \(U^{(0)}\), reaching local maxima of \(\rho = 0.87\) and \(\rho = 0.86\), respectively, at an ensemble size of eleven (the red dashed circle in Fig.~\ref{Fig:2_UQ_on_OMat24}\textbf{b}). This underscores the effectiveness of inverse‑RMSE weighting in suppressing noise from lower‑accuracy members. The eleven models included in the optimal ensemble are highlighted by the red dashed box in Fig.~\ref{Fig:2_UQ_on_OMat24}\textbf{a}. Notably, \(\rho\) for \(U^{(1)}\) decreases slightly up to fourteen models and then increases as additional lower‑accuracy models are added, demonstrating that the diversity contributed by less accurate members can also enhance performance. These findings underscore the value of harnessing the architectural diversity of existing uMLIP models to improve UQ. In comparison with \(U_2\), \(U_1\) consistently outperforms it. Therefore, we propose the \(U^{(1)}\) metric, computed from an ensemble of eleven uMLIPs, as a universal uncertainty metric for general inorganic materials, hereafter denoted \(U\).
The weights for each model are shown in Table~\ref{tab:uMLIP}. 

Fig.~\ref{Fig:2_UQ_on_OMat24}\textbf{c} shows a hexbin parity plot of the predicted uncertainty \(U\) against the actual force error on the OMat24 test set. The density of points closely follows the ideal \(y=x\) (orange dashed line), with Spearman’s \(\rho=0.87\), indicating a strong monotonic relationship between uncertainty and error. Notably, the conditional spread around the diagonal remains under one order of magnitude across nearly five decades of \(U\) (\(10^{-3}\)–\(10^{2}\)\,eV/\AA), indicating that low‑uncertainty predictions almost never produce large errors, while high‑uncertainty cases reliably signal catastrophic deviations. This tight, nearly unbiased clustering demonstrates that \(U\) directly corresponds to force error without the need for post hoc calibration.

Fig.~\ref{Fig:2_UQ_on_OMat24}\textbf{d} shows a hexbin plot of the Orb‐confidence against the true force error on the OMat24 test set~\cite{rhodes2025orb}, directly comparable to the ensemble‐based $U$ in Fig.~\ref{Fig:2_UQ_on_OMat24}\textbf{c}. While both metrics achieve a Spearman’s $\rho=0.87$, Orb‐confidence exhibits a much narrower horizontal spread (only $\sim$2-3 decades of confidence values) and a large vertical dispersion: at a single confidence level, the force error can vary by up to two orders of magnitude. In particular, some configurations labeled with moderate confidence (10-20) still show catastrophic errors ($>$10 eV/\AA), indicating that Orb‐confidence cannot reliably flag its worst failures.
This improved calibration of $U$ relative to Orb‐confidence translates into tangible gains, as shown by the accuracy-coverage curves for total energies and atomic forces (Fig.~\ref{Fig:2_UQ_on_OMat24}\textbf{e},\textbf{f}). For $U$, all RMSE values are computed using the ensemble mean of an eleven-member uMLIP, denoted $\langle\mathrm{uMLIP}\rangle$. For Orb-confidence, RMSE values are calculated solely by Orb. 
Up to approximately \(80\%\) coverage, \(U\) maintains the energy RMSE below \(0.05\,\mathrm{eV/atom}\) and the force RMSE below \(0.06\,\mathrm{eV/\AA}\). In sharp contrast, Orb‐confidence can only achieve the same accuracy below roughly \(25\%\) coverage for energy and \(40\%\) for force. Beyond these thresholds, its error rises rapidly, particularly when the most challenging \(\sim10\%\) of configurations are included (around \(90\%\) coverage), highlighting the substantial advantage of \(U\) in identifying high‑error cases.

Building on our analysis of uncertainty calibration, we next consider whether to use the ensemble mean or a single top-performing model to replace DFT. In principle, averaging a homogeneous ensemble can cancel random noise and improve accuracy, but our uMLIP ensemble is heterogeneous, so this effect may not hold. Accordingly, we compare accuracy-coverage curves computed with $\langle\mathrm{uMLIP}\rangle$ against those obtained using the single most accurate model, eqV2-31M-omat. Fig.~\ref{Fig:2_UQ_on_OMat24}\textbf{e},\textbf{f} show that eqV2-31M-omat matches or outperforms the ensemble mean at nearly every coverage level, maintaining lower RMSE values. Accordingly, we adopt eqV2-31M-omat as the surrogate for DFT reference to calculate forces and eqV2-31M-OAM to calculate energy in all subsequent sections.

\subsection*{Validation of $U$ across diverse materials}
To further demonstrate the universality of our uncertainty metric $U$, we apply it to an extensive collection of PBE‑level DFT datasets employed in previous sMLIP development, encompassing a broad spectrum of materials. Further details of the datasets are provided in Supplementary Note 1 and Table~\ref{tab:DFT-valid}. These datasets are classified into three categories according to their elemental composition and structural motifs:
\begin{enumerate}
    \item \textbf{Metals and Alloys}: Pure metals (e.g., Fe, Mg), the full set of transition metals (TM23), and medium‑ to high‑entropy alloys such as CrCoNi, VCoNi, MoNbTaVW, MoNbTaTi, WTaCrV, and the M16 binary alloys (264,383 configurations, 13,701,879 atoms).
    \item \textbf{Inorganic Compounds}: Compounds and interstitial solid solutions involving light elements (H, C, N, O) combined with metals, including FeH, LiH, FeC, MoNbTaWH, HfO$_2$, Ga$_2$O$_3$, and GaN (49,092 configurations, 4,412,916 atoms).
    \item \textbf{Other Materials}: Carbon, metal–organic frameworks (MOFs), ionic covalent organic frameworks (ICOFs), surface‑catalytic structures, perovskites, and battery‑relevant materials such as LiPS, Li$_4$P$_2$O$_7$, and various cathode compositions (64,464 configurations, 4,868,199 atoms).
\end{enumerate}

\begin{figure}
    \centering
    \includegraphics[width=1\linewidth]{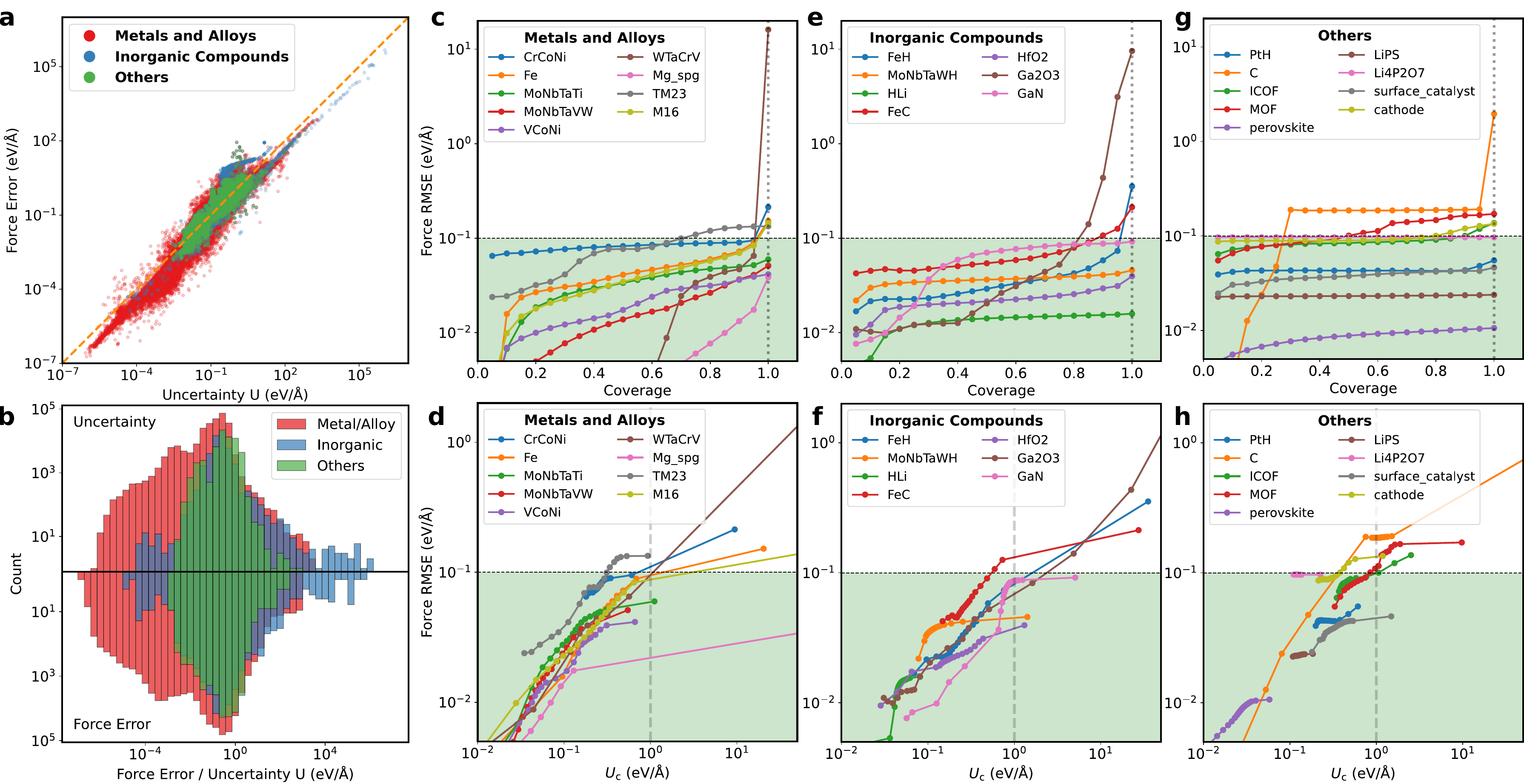}
    \caption{
\textbf{Performance of the uncertainty indicator \( U \) on additional datasets.} 
\textbf{a} Scatter plot of uncertainty \( U \) versus force error across various datasets. The data are grouped into three categories based on their origin: metals and alloys, inorganic compounds, and others (including MOFs, perovskites, etc.). The distribution is centered along the diagonal \( y = x \), indicating a strong correlation between uncertainty and error.
\textbf{b} Histograms of \( U \) and force error, shown above and below the x-axis, respectively, for the three data categories. The distributions of uncertainty and error closely resemble each other within each group.
\textbf{c}-\textbf{h} For each dataset, configurations with U higher than uncertainty criteria \( U_c \) are gradually removed, and the RMSE of the remaining “low-uncertainty” configurations is plotted as a function of dataset coverage (c, e, g) and \( U_c \) (d, f, h).
}
\label{Fig:3_UQ_on_other_dataset}
\end{figure}

Fig.~\ref{Fig:3_UQ_on_other_dataset}\textbf{a} plots the predicted uncertainty \(U\) against the true force error for each of the three dataset categories. Compared to the OMat24 test set (Fig.~\ref{Fig:2_UQ_on_OMat24}\textbf{c}), both \(U\) and the force error now span an even broader range (\(10^{-7}\)–\(10^{6}\)\,eV/\AA). Nevertheless, a strong monotonic relationship persists: Spearman’s \(\rho\) is 0.92 for metals and alloys, 0.88 for inorganic compounds, and 0.82 for the remaining materials, demonstrating that higher \(U\) values reliably correspond to larger errors across all categories. 
Notably, the metals and alloys attain a higher correlation than the OMat24 benchmark (\(\rho = 0.87\); Fig.~\ref{Fig:2_UQ_on_OMat24}\textbf{c}), underscoring the robustness of \(U\) in comparatively uniform systems and contrasting with the greater heterogeneity of the other two categories.
Fig.~\ref{Fig:3_UQ_on_other_dataset}\textbf{b} shows mirror‐histograms of the predicted uncertainty \(U\) (top) and the actual force error (bottom) for each material category. The close symmetry of each color‐coded distribution around the horizontal axis demonstrates that \(U\) faithfully captures the error spread across all systems. Moreover, the histograms reveal distinct accuracy regimes: metals and alloys (red) concentrate almost entirely below \(10^{0}\)\,eV/\AA, indicating relatively high accuracy; inorganic compounds (blue) span a wider range and dominate above \(10^{0}\), reflecting greater variability; the “others” group (green) occupies an intermediate region around \(10^{-2}\)–\(10^{0}\). These differences underscore that, beyond uncertainty, the baseline uMLIP accuracy varies significantly by material class, with metals and alloys achieving the best performance, consistent with recent benchmarking studies~\cite{shuang2025universal}. 

Fig.~\ref{Fig:3_UQ_on_other_dataset}\textbf{c}, \textbf{e} and \textbf{g} display coverage-accuracy curves in which we progressively discard the highest‐uncertainty configurations and record the resulting force RMSE on the remaining data. The shaded green region denotes RMSE $<$ 0.1\,eV/\AA, a common requirement for developing reliable sMLIPs. For every material class, removing even a small fraction of the most uncertain points produces a clear, monotonic drop in error. The most dramatic improvements occur in systems with large initial RMSE: in the W-Ta-Cr-V alloy (brown line in Fig.~\ref{Fig:3_UQ_on_other_dataset}\textbf{c}), discarding under 5\% of configurations brings the RMSE down from 10\,eV/\AA \ to below 0.1\,eV/\AA. Similarly, steep declines are seen for CrCoNi, Ga$_2$O$_3$, FeH and FeC, confirming that uMLIP yields high‐accuracy predictions for the vast majority of structures, with only a handful of OOD points driving the worst errors. Datasets whose baseline RMSE already lies inside the green region show nearly flat curves, indicating uniformly reliable performance of the uMLIP. Dataset Carbon (C) is an exception, as its RMSE remains elevated until more than 70\% of the data are retained, suggesting that uncertainty and error are less tightly coupled in this chemically simple but structurally varied system.

Practical workflows often lack access to true DFT errors and must instead rely on an a priori uncertainty cutoff \(U_c\) to flag unreliable uMLIP predictions. 
Fig.~\ref{Fig:3_UQ_on_other_dataset}\textbf{d}, \textbf{f}, and \textbf{h} show how the force RMSE of the retained subset varies as a function of \(U_c\).
The point where each curve intersects the horizontal target line at \(\mathrm{RMSE} = 0.1\,\mathrm{eV/\AA}\) defines the \(U_c\) threshold below which uMLIP predictions can be trusted and catastrophic errors on OOD configurations can be avoided.
For both metals and alloys (Fig.~\ref{Fig:3_UQ_on_other_dataset}\textbf{d}) and inorganic compounds (Fig.~\ref{Fig:3_UQ_on_other_dataset}\textbf{f}), these intersection points fall at or above \(U_c = 1\,\mathrm{eV/\AA}\), with only a few outliers (e.g. TM23, FeC) requiring slightly lower $U_c$. In the “Others” category (Fig.~\ref{Fig:3_UQ_on_other_dataset}\textbf{h}), which contains a larger fraction of OOD configurations, the required cutoff shifts marginally below \(U_c = 1\,\mathrm{eV/\AA}\), indicating that dataset‑specific tuning may improve performance in these more challenging regimes. Given the broad compositional and structural diversity tested, the consistency of this threshold is striking. We therefore recommend \(U_c = 1\,\mathrm{eV/\AA}\) as a general rule of thumb for selecting configurations that uMLIP can predict with RMSE \(\le 0.1\,\mathrm{eV/\AA}\). Users may adjust this cutoff to meet more stringent or relaxed accuracy requirements; for example, setting \(U_c = 0.3\,\mathrm{eV/\AA}\) yields an RMSE of approximately 0.05\,eV/\AA.

\subsection*{Uncertainty-aware model distillation for W}
Building on the demonstrated efficacy of \(U\) for quantifying uncertainty in uMLIP ensembles, we now turn to its most direct application: guiding the distillation of predictive accuracy into streamlined sMLIP models. In this section, we introduce an uncertainty-aware model distillation (UAMD) framework that leverages \(U\) to adaptively construct the training dataset by retaining low-\(U\) configurations using the most accurate predictions in place of expensive DFT references, and by flagging high-\(U\) configurations for targeted DFT calculations. As discussed below, UAMD drastically reduces and in some cases entirely eliminates the need for expensive DFT calculations compared to conventional sMLIP development workflows. Furthermore, unlike standard model distillation techniques, UAMD uses uncertainty estimates to explicitly control and constrain error propagation from the uMLIP ensemble, ensuring that the resulting sMLIP meets its target accuracy with a minimal amount of reference data. 

To demonstrate the UAMD framework, we first consider the tungsten (W) system as a case study. The primary configuration set comprises 1{,}026 structures from Ref.~\cite{shuang2025modeling}, covering a broad spectrum of defects. To extend this set into the extreme deformation regime relevant to radiation damage, we augment it with 13 dimer and 100 short‑range configurations from Byggmästar et al.~\cite{byggmastar2019machine}. These highly extreme structures with huge forces lie outside the standard uMLIP training domain and thus serve as challenging OOD test cases for UAMD. 

\begin{figure}
    \centering
    \includegraphics[width=1\linewidth]{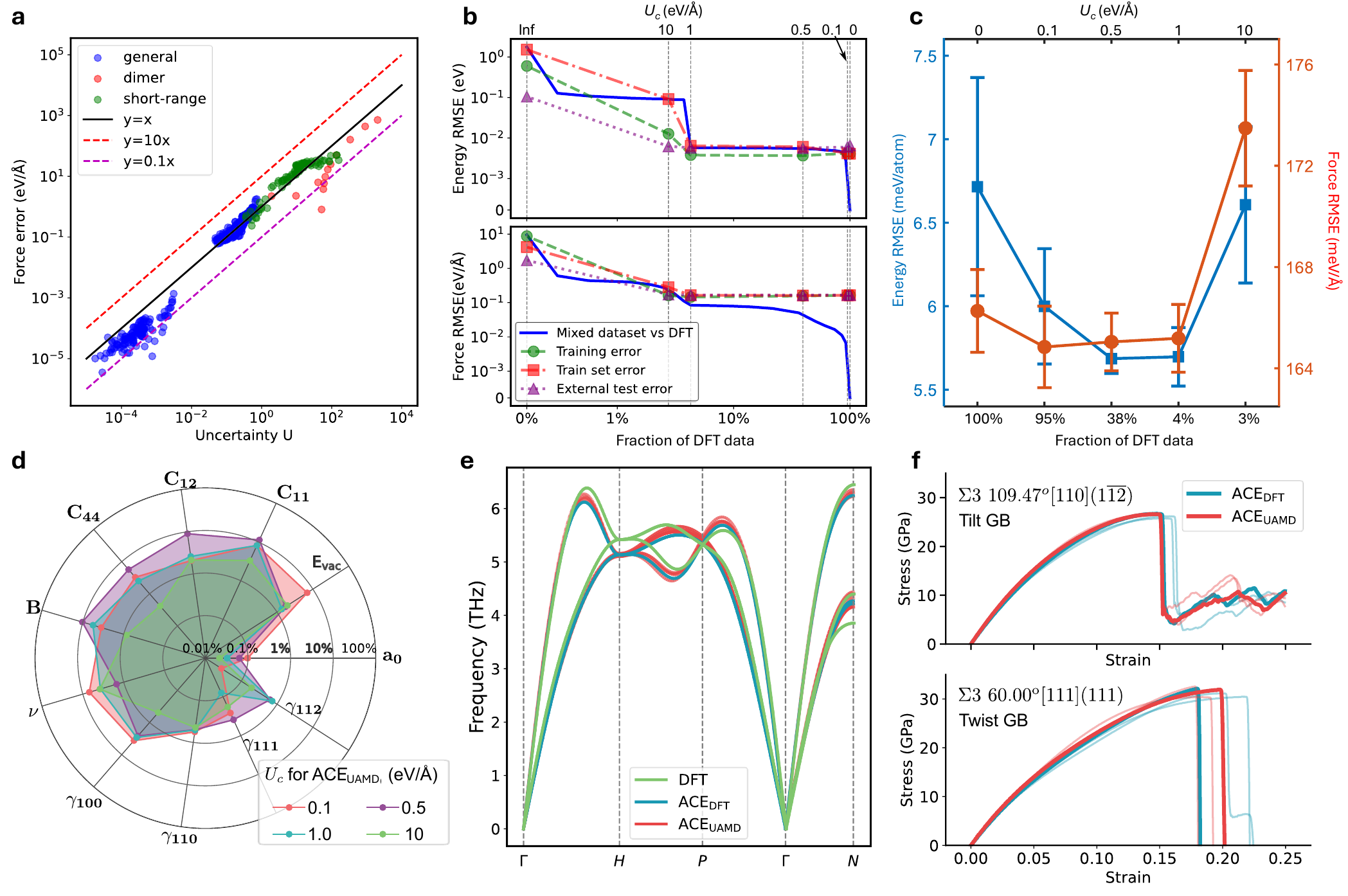}
    \caption{\textbf{Validation of uncertainty-aware model distillation for W.} 
    \textbf{a} The uncertainty \( U \) shows a strong correlation with force errors across 1{,}139 configurations of W. Dimer and short-range configurations (in red and green, respectively) exhibit both high uncertainty and high errors.
\textbf{b} Hybrid datasets are constructed using different uncertainty thresholds \( U_c \): configurations with \( U < U_c \) use UMLIP predictions, while those with \( U \geq U_c \) are labeled using DFT. The x-axis indicates the fraction of DFT data. Blue solid and green dashed lines represent the RMSE of the dataset and ACE training error, respectively. Red dash-dot and purple dotted lines denote ACE performance on training and test sets, respectively.
\textbf{c} Basic physical properties predicted by the ACE model trained on the hybrid dataset.
\textbf{d-e} Comparison of phonon dispersion curves and stacking fault energy profiles predicted by ACE and DFT.
\textbf{f} Stress-strain curves from uniaxial tension simulations of \(\Sigma 3 \) tilt (above) and twist (below) grain boundaries. ACE$_\text{DFT}$ and ACE$_\text{mix}$ models produce nearly identical results.}
    \label{Fig:4_W_system}
\end{figure}

Fig.~\ref{Fig:4_W_system}\textbf{a} plots the predicted uncertainty \(U\) against the true force error for the W configurations. The dimer and short‐range points lie predominantly in the high‐\(U\), high‐error quadrant, confirming that these OOD geometries are correctly identified as challenging. In contrast, the bulk of the general dataset clusters at low \(U\) and low error, with many samples below \(10^{-2}\,\mathrm{eV/\AA}\). Despite spanning over seven orders of magnitude in \(U\) (\(10^{-4}\)–\(10^{3}\,\mathrm{eV/\AA}\)), the points follow the \(y=x\) diagonal closely, with only a few outliers dropping below the \(y=0.1x\) reference lines. This tight, nearly linear trend demonstrates that \(U\) provides a robust, physically meaningful measure of prediction error even for extreme deformations.  

We then apply a series of uncertainty thresholds \(U_c\) to partition the dataset into high‑ and low‑$U$ subsets. Configurations with \(U>U_c\) are assigned true DFT labels, while those with \(U\le U_c\) use uMLIP predictions. For consistency with prior benchmarks~\cite{shuang2025universal}, we employ eqV2‑31M‑OAM for energies and eqV2‑31M‑omat for forces. We evaluate six thresholds: \(U_c=\infty\) (i.e.\ trust uMLIP for all points), 10, 1, 0.5, 0.1, and 0 (i.e.\ full DFT). Higher \(U_c\) values correspond to more permissive use of uMLIP labels, whereas lower values increasingly rely on the DFT reference.
For each mixed dataset, we train an ACE potential (see Methods) and report the resulting errors in Fig.~\ref{Fig:4_W_system}\textbf{b}. The fraction of DFT‐labeled configurations (measured by atomic environments) is indicated along the bottom of the x‐axis, while the corresponding \(U_c\) values are shown on the top x‑axis.
The blue solid line shows the error of the constructed mixed dataset relative to DFT values. Incorporating only about 4\% of the configurations as DFT references reduces the energy RMSE from over 1~eV to below 0.01~eV and the force RMSE from more than 10~eV/\AA{} to approximately 0.1~eV/\AA. These remaining errors reflect the propagation of data error from the uMLIP predictions into the distilled sMLIP.
The green dashed line shows the ACE training error, defined as the RMSE between ACE predictions and the mixed dataset. Introducing approximately 4\% of DFT references leads a sharp drop in loss, and by 4\% the training error stabilizes. The relatively high training error for the full uMLIP data arises from the non‑conservative nature of the eqV2 models, in which forces are not exact derivatives of the predicted energy. The plateau in training error beyond 4\% reflects the irreducible errors inherent to the ACE training process.
Furthermore, we evaluate each ACE potential on both the full DFT dataset and an independent validation set that includes complex plasticity and crack propagation scenarios, as shown in Fig.~\ref{FigS-npj-test} and sourced from Ref.~\cite{shuang2025modeling}, to assess their generalization performance. Fig.~\ref{Fig:4_W_system}\textbf{b} shows that once \(U_{c}<1\) (i.e.\ when about 4\% of the configurations are labeled with DFT), the errors on the full DFT set and on the independent test set converge to nearly the same value. This pattern demonstrates that the final accuracy of the ACE model is determined by two primary contributions: the intrinsic data error of the training labels (blue solid line) and the training error during fitting (green dashed line).
Importantly, the observed test error on the independent DFT dataset is not simply the sum of the training‐data error and the fitting error; rather, it is governed by whichever contribution is dominant, and the two can partially offset each other. In energy predictions, the test error primarily reflects the training‐data error, whereas in force predictions, where the fitting error is larger, the test error closely follows the training‐loss curve. 

To investigate the impact of the uncertainty threshold \(U_c\), and thus the fraction of DFT‑labeled data, on model performance, we plot energy and force RMSEs for \(\mathrm{ACE}_{\mathrm{UAMD}}\) trained with different \(U_c\) values in Fig.~\ref{Fig:4_W_system}\textbf{c}. Fig.~\ref{Fig:4_W_system}\textbf{a} shows that raising \(U_c\) (i.e., trusting more uMLIP‑generated labels) admits additional high‑error configurations, which might naively be expected to worsen accuracy. Instead, Fig.~\ref{Fig:4_W_system}\textbf{c} reveals that the lowest RMSEs occur at intermediate DFT fractions (\(\sim 4\%\) to \(39\%\)), not with either full DFT or full uMLIP labels. This behavior reflects a trade‑off between label bias and label noise. DFT labels provide low‑bias physical information yet carry stochastic numerical noise, arising from finite plane‑wave cutoffs, \(k\)‑point sampling, and incomplete convergence~\cite{DFT-error-JPCA,DFT-error-npj}. In contrast, uMLIP predictions are smooth and effectively free of the non‑physical numerical fluctuations, but they exhibit systematic bias in regions of large model error. An optimal mixture corrects the uMLIP bias while limiting exposure to DFT‑induced noise: below \(\approx 4\%\) DFT coverage residual uMLIP bias dominates, whereas above \(\approx 39\%\) the increasing DFT noise and overfitting begin to degrade performance. Selecting \(U_c=1\,\mathrm{eV/\AA}\) (yielding about \(4\%\) DFT labels) achieves this balance, reducing the number of expensive DFT calculations while improving fidelity beyond conventional fully DFT‑labeled training.

While Fig.~\ref{Fig:4_W_system}\textbf{b} and \textbf{c} presents RMSE comparisons between ACE and DFT, it is essential to validate the accuracy of ACE\textsubscript{UAMD} in predicting physical properties at different scales, comparing to ACE\textsubscript{DFT} model, which uses fully DFT-labeled data. Fig.~\ref{Fig:4_W_system}\textbf{d} shows the relative errors on several basic properties of W with various $U_{c}$. For each \(U_c\), we independently trained five ACE models. The markers denote the mean predictions over these models. The lattice constant \(a\) remains exceptionally stable across all \(U_c\), with relative deviations below 0.1\%. Surface energies \(\gamma_{100}\) and vacancy formation energy \(E_{\mathrm{vac}}\) also show minor sensitivity, with relative errors around 3\% and other surface energies under 1\%. By contrast, elastic properties such as \(C_{11}\), \(C_{12}\), \(C_{44}\), bulk modulus \(B\), and Poisson’s ratio \(\nu\) display greater dependence on \(U_c\), with relative errors approaching 10\% at \(U_c = 1\). This increased sensitivity likely arises because elastic constants are determined by derivatives of energy and stress near equilibrium, making them more susceptible to small training data inconsistencies. 

Fig.~\ref{Fig:4_W_system}\textbf{e} compare the phonon spectrums predicted by ACE\textsubscript{UAMD} and ACE\textsubscript{DFT} against reference DFT data~\cite{pan2024atomic,xu2020frank}. Both ACE variants produce nearly identical curves that track the DFT results closely, with a slight systematic underestimation of vibrational frequencies. The close agreement between ACE\textsubscript{UAMD} and ACE\textsubscript{DFT} indicates that adding more high‐accuracy DFT data beyond the UAMD threshold does not yield significant gains for these properties, suggesting diminishing returns for further refinement in these regimes.
Fig.~\ref{Fig:4_W_system}\textbf{f} shows stress-strain curves of bicrystal tensile molecular dynamics (MD) simulation using the ACE\textsubscript{UAMD} potential (with \(U_c = 1.0\,\mathrm{eV/\AA}\)) and the ACE\textsubscript{DFT} reference. Two \(\Sigma 3\) grain boundary (GB) geometries are studied: a symmetric tilt boundary and a twist boundary, see Methods. Each loading simulation is repeated three times with different random velocity seeds to confirm reproducibility. The details of deformation mechanisms are shown in Fig.~\ref{FigS-W-deformation}. The ACE\textsubscript{UAMD} model faithfully captures the contrasting deformation modes seen with the ACE\textsubscript{DFT} potential, namely localized plastic slip at the tilt boundary and brittle cleavage along the twist boundary. Both the yield stresses and the strains at which plastic flow or fracture initiate agree closely between the two potentials, demonstrating that UAMD‐trained potentials can reliably reproduce complex mechanical responses. 

Taken together, the W case study demonstrates that an ACE potential trained with only 4\% of configurations labeled by DFT (\(U_c = 1.0\,\mathrm{eV/\AA}\)) matches the accuracy of fully DFT‐trained models across energetic, structural, and mechanical benchmarks.

\subsection*{Uncertainty-aware model distillation for MoNbTaW alloys}

We next apply UAMD to develop sMLIPs for the prototypical refractory high‐entropy alloy MoNbTaW. Conventional sMLIPs for HEAs are typically trained on narrow composition ranges and include only limited defect types in pure metals, leaving them unable to capture the full chemical and structural complexity of real microstructures~\cite{HEA-SNAP}. In principle, a truly general‐purpose HEA potential would require an exhaustive DFT dataset that samples every composition, chemical interaction, and defect configuration, a requirement that becomes infeasible as each additional element exponentially expands the combinatorial space and the associated DFT workload. UAMD promises to overcome this bottleneck by drastically reducing the number of required DFT calculations.

\begin{figure}
    \centering
    \includegraphics[width=1\linewidth]{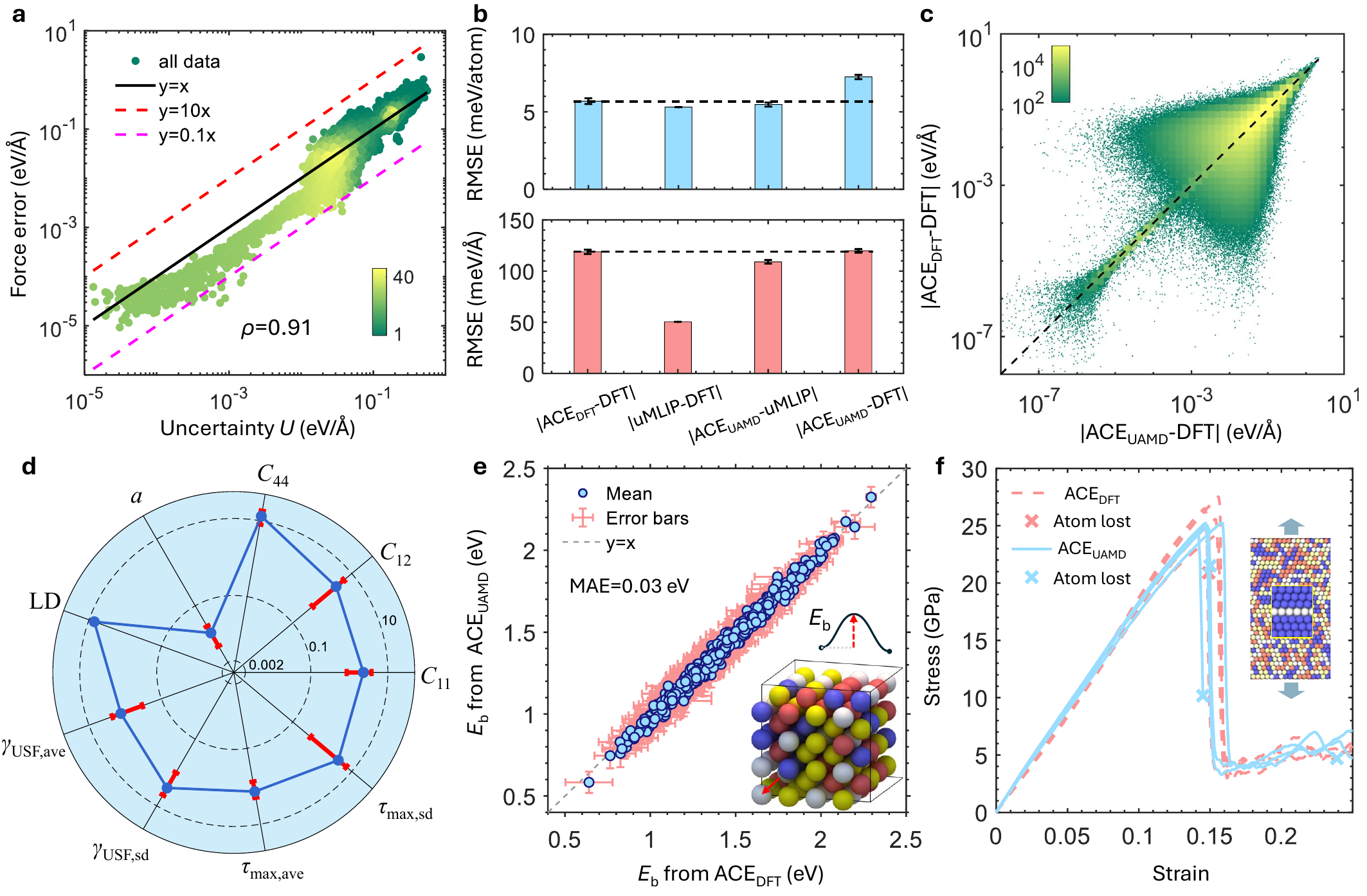}
    \caption{\textbf{Validation of uncertainty-aware model distillation in MoNbTaW HEA systems.}
\textbf{a} Correlation between force error and predicted uncertainty; each point represents a configuration.
\textbf{b} Energy (top) and force (bottom) RMSEs across various scenarios. The dashed lines indicate the RMSEs of ACE models trained solely on DFT data.
\textbf{c} Correlation of per-atom errors between ACE models trained on DFT data and those trained via UAMD, both evaluated against raw DFT references. Each point represents an atom.
\textbf{d} Comparison of various mechanical properties predicted by $\text{ACE}\text{DFT}$ and $\text{ACE}\text{UAMD}$.
\textbf{e} Validation of monovacancy diffusion barriers ($E_b$) in Mo\textsubscript{25}Nb\textsubscript{25}Ta\textsubscript{25}W\textsubscript{25}, based on 500 independent NEB calculations with randomly shuffled atomic positions. Error bars reflect the spread from five independently trained ACE models.
\textbf{f} Validation in bicrystal tensile simulations. For both DFT- and UAMD-derived ACE models, five different potentials are tested. Cross markers indicate simulation failure due to atom loss.}
    \label{Fig:5_MoNbTaW}
\end{figure}

We benchmark UAMD on a curated dataset of 17{,}654 MoNbTaW configurations, which includes ground‑state structures, finite elastic strains, ab initio molecular dynamics (AIMD) snapshots, and a wide variety of point and extended defects across the full compositional space~\cite{IJP-HEA-MLIP}. Similar to the workflow applied to OMat24 and W, we first examine the relationship between \(U\) and the force prediction error. As shown in Fig.~\ref{Fig:5_MoNbTaW}\textbf{a}, a strong monotonic correlation is observed, with Spearman’s \(\rho = 0.91\).
High values of \(U\) reliably identify configurations with large force errors, while the majority of points occupy the low‑\(U\), low‑error region (\(<10^{-1}\,\mathrm{eV/\AA}\)). Importantly, no configuration exceeds the critical threshold \(U_c = 1\,\mathrm{eV/\AA}\), and both the force errors and uncertainties remain small for the overwhelming majority of data. This robust performance allows us to replace DFT labels with uMLIP predictions for all configurations (eqV2‑31M‑OAM model for energy and eqV2‑31M‑omat for force), without any DFT calculation. These surrogate labels form the training data for \(\mathrm{ACE}_{\mathrm{UAMD}}\). For comparison, we also train \(\mathrm{ACE}_{\mathrm{DFT}}\) on the full DFT dataset. To ensure statistical robustness, each ACE model variant is trained five times with different random initializations.

The performance of the resulting models is quantified by energy and force RMSEs, as shown in Fig.~\ref{Fig:5_MoNbTaW}\textbf{b}. We compare four scenarios: (i) \(\mathrm{ACE}_{\mathrm{DFT}}\) predictions against DFT references; (ii) uMLIP predictions against DFT references; (iii) \(\mathrm{ACE}_{\mathrm{UAMD}}\) predictions against uMLIP outputs; and (iv) \(\mathrm{ACE}_{\mathrm{uMLIP}}\) predictions against DFT references. 
We observe three key findings. First, the uMLIPs used for data distillation outperform \(\mathrm{ACE}_{\mathrm{DFT}}\) in both energy and force prediction, which is consistent with our recent study~\cite{shuang2025universal}. Second, \(\mathrm{ACE}_{\mathrm{UAMD}}\) aligns more closely with the uMLIP predictions than \(\mathrm{ACE}_{\mathrm{DFT}}\) does with DFT, suggesting that uMLIP outputs provide a smoother target that reduces numerical noise and simplifies ACE training. Third, when evaluated against true DFT data, \(\mathrm{ACE}_{\mathrm{UAMD}}\) exhibits only slightly higher energy RMSE (7.25 vs. 5.69 meV/atom) and maintains a force RMSE comparable to that of \(\mathrm{ACE}_{\mathrm{DFT}}\) (118.82 meV/\AA{} vs. 119.83 meV/\AA). These results confirm that error propagation from the uMLIP teacher to the ACE student remains well controlled even in the absence of DFT labels.

Fig.~\ref{Fig:5_MoNbTaW}\textbf{c} shows the density of per‐atom force errors for \(\mathrm{ACE}_{\mathrm{DFT}}\) versus \(\mathrm{ACE}_{\mathrm{UAMD}}\) on a logarithmic scale, with both models evaluated against the same DFT reference. The pronounced symmetry about the \(y=x\) diagonal indicates that neither potential holds a systematic advantage: configurations where \(\mathrm{ACE}_{\mathrm{UAMD}}\) has larger error than \(\mathrm{ACE}_{\mathrm{DFT}}\) are balanced by the opposite case. Similar trends are observed in Fig.~\ref{FigS-UAMD-HEA-error}b,c when comparing the absolute errors \(\lvert \mathrm{ACE}_{\mathrm{DFT}}-\mathrm{DFT}\rvert\) vs. \(\lvert \mathrm{uMLIP}-\mathrm{DFT}\rvert\) and \(\lvert \mathrm{ACE}_{\mathrm{UAMD}}-\mathrm{uMLIP}\rvert\) vs. \(\lvert \mathrm{uMLIP}-\mathrm{DFT}\rvert\), indicating that the random‑error distributions are similar across these sources. Formally, the error of \(\mathrm{ACE}_{\mathrm{UAMD}}\) decomposes into its residual fitting error to the uMLIP teacher (training error) plus the teacher’s zero‐mean numerical noise from DFT (data error, as shown in Fig.~\ref{FigS-UAMD-HEA-error}a). In contrast, \(\mathrm{ACE}_{\mathrm{DFT}}\) fits the raw DFT labels directly under identical model capacity and regularization, making it susceptible to numerical errors. By training on the inherently smoother uMLIP ensemble predictions, \(\mathrm{ACE}_{\mathrm{UAMD}}\) effectively filters out these DFT fluctuations, much like deliberate noise injection, which is known to suppress overfitting and improve generalization~\cite{NC-noise-MLIP}. In regions where the opposite interplay occurs, the two methods counterbalance one another, yielding statistically indistinguishable fidelity to DFT forces across the , as illustrated in Fig.\ref{Fig:5_MoNbTaW}\textbf{b}. These results are consistent with the case for W in Fig.~\ref{Fig:4_W_system}.

To further validate the efficacy of the distilled model, we use \(\mathrm{ACE}_{\mathrm{UAMD}}\) to predict key properties of \(\mathrm{Mo}_{25}\mathrm{Nb}_{25}\mathrm{Ta}_{25}\mathrm{W}_{25}\), including the lattice constant \(a\), lattice distortion (LD), independent elastic constants \(C_{11}\), \(C_{12}\), \(C_{44}\), and the statistics of unstable stacking fault energies (\(\gamma_{\mathrm{USF,ave}}\), \(\gamma_{\mathrm{USF,std}}\)) and maximum restoring forces (\(\tau_{\mathrm{USF,ave}}\), \(\tau_{\mathrm{USF,std}}\)), which are critical to determine the solid solution strengthening of HEAs~\cite{Acta-HEA-shuang}. As shown in Fig.~\ref{Fig:5_MoNbTaW}\textbf{d}, all of these quantities agree closely with the corresponding \(\mathrm{ACE}_{\mathrm{DFT}}\) predictions, exhibiting only minor deviations. We then benchmark monovacancy migration barriers (Fig.~\ref{Fig:5_MoNbTaW}\textbf{e}), where \(\mathrm{ACE}_{\mathrm{UAMD}}\) and \(\mathrm{ACE}_{\mathrm{DFT}}\) produce virtually identical energy profiles, and perform bicrystal tensile simulations (Fig.~\ref{Fig:5_MoNbTaW}\textbf{f}), which reveal matching stress–strain responses from the elastic regime through yield and into plastic flow.

\subsection*{General-purpose potential for MoNbTaW alloys}
Although the 17{,}654‑configuration MoNbTaW dataset provides a rigorous benchmark, it does not span the full spectrum of microstructural motifs required for general‑purpose modelling, including arbitrary GBs, dislocation–GB interactions, and fracture. As illustrated in Fig.~\ref{Fig:5_MoNbTaW}\textbf{f}, MD simulations of bicrystal tension can terminate prematurely with the LAMMPS "Atom lost" error. The strength of the UAMD framework is its capacity to explore vast configurational spaces without incurring the cost of additional DFT calculations. To demonstrate this capability, we begin with the recently published defect genome for W~\cite{shuang2025modeling}, which systematically samples general plasticity, surfaces, and crack tips. We then apply the maximum‑volume algorithm in the MLIP‑2 package~\cite{MLIP-2} to introduce increasing chemical complexity into each topological motif across the full MoNbTaW compositional space (see Supplementary Note~2). In total, we generate 7{,}000 defect configurations, providing broad coverage for general‑purpose modelling of MoNbTaW alloys.

\begin{figure}
    \centering
    \includegraphics[width=1\linewidth]{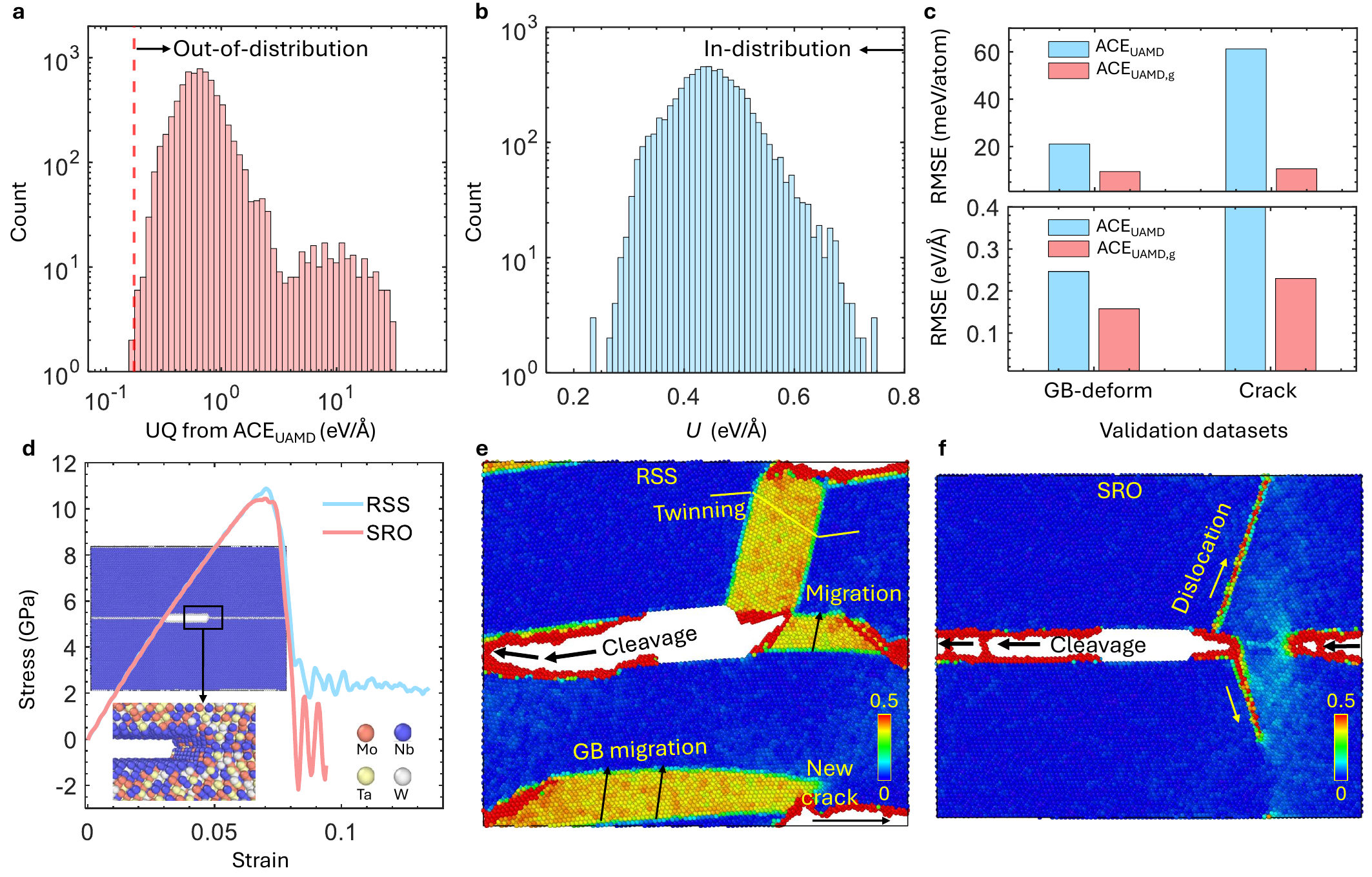}
\caption{\textbf{Application of UAMD-derived general-purpose potential in modeling high-entropy MoNbTaW alloys.} 
\textbf{(a)} Uncertainty quantification distribution using conventional ensemble learning based on $\text{ACE}_{\text{UAMD}}$ models for 7,000 newly generated configurations. 
\textbf{(b)} Distribution of the new uncertainty metric $U$ for the same dataset. 
\textbf{(c)} Performance comparison between $\text{ACE}_{\text{UAMD}}$ and $\text{ACE}_{\text{UAMD,g}}$ on independent test datasets. 
\textbf{(d)} Stress-strain curves of bicrystal fracture for random solid solution (RSS) and chemically short-range ordered (SRO) systems. The inset shows the bicrystal model with a center crack, highlighting Nb segregation at both the crack surface and grain boundary (GB). 
\textbf{(e,f)} Per-atom atomic shear strain in final snapshots for RSS and SRO cases, highlighting critical deformation mechanisms including deformation twinning, cleavage, GB migration, and dislocation emission.}
\label{Fig:potential_application}
    \label{Fig:6_MoNbTaW_append_data}
\end{figure}

We first use the original \(\mathrm{ACE}_{\mathrm{UAMD}}\) ensemble, trained on 17{,}654 configurations, to evaluate the 7{,}000 newly generated structures. Fig.~\ref{Fig:6_MoNbTaW_append_data}\textbf{a} shows that all of these configurations lie outside the distribution spanned by the original dataset. We then compute the uncertainty \(U\) for each new configuration and find that every value falls below the threshold \(U_{c} = 1\,\mathrm{eV/\AA}\) (Fig.~\ref{Fig:6_MoNbTaW_append_data}\textbf{b}), confirming that the uMLIP predictions remain reliable. We merge these configurations with the original dataset, and assign uMLIP labels to all 24{,}654 configurations. From this augmented set, we train a general-purpose ACE potential, denoted \(\mathrm{ACE}_{\mathrm{UAMD,g}}\). We then benchmark both \(\mathrm{ACE}_{\mathrm{UAMD,g}}\) and the original \(\mathrm{ACE}_{\mathrm{UAMD}}\) on two independent DFT test sets: GB‑deform, which probes severe GB deformations, and fracture, which examines crack propagation, across varying compositions. Remarkably, \(\mathrm{ACE}_{\mathrm{UAMD,g}}\) achieves substantially lower errors on both test sets compared to \(\mathrm{ACE}_{\mathrm{UAMD}}\) as shown in Fig.~\ref{Fig:6_MoNbTaW_append_data}\textbf{c}, despite relying solely on uMLIP‑generated labels. The accuracy gains for GB‑deform and fracture approach those obtained for W using fully DFT‑labeled defect genomes~\cite{shuang2025modeling}, demonstrating the capability of UAMD to extend sMLIPs efficiently to new regions of configuration space without costly DFT calculations. 

Finally, we apply the \(\mathrm{ACE}_{\mathrm{UAMD,g}}\) potential to large‑scale MD simulations of bicrystal fracture in MoNbTaW alloys, capturing coupled chemical and mechanical effects on crack propagation. We first use hybrid Monte Carlo/molecular dynamics (MC/MD) to probe chemical short‑range order (SRO) in a bicrystal model containing a central crack. After equilibration, the inset of Fig.~\ref{Fig:6_MoNbTaW_append_data}\textbf{d} shows pronounced segregation of Nb atoms to the GB and crack surfaces. The details of SRO formation are presented in Fig.~\ref{FigS-HEA-general}a-d. The stress-strain curves in Fig.~\ref{Fig:6_MoNbTaW_append_data}\textbf{d} compare the mechanical response of bicrystals with a random solid solution (RSS) and with SRO. We find that SRO has little influence on the elastic regime, slightly reduces the yield stress, and substantially diminishes ductility, with loss of load‑bearing capacity occurring at only 8\% tensile strain. Fig.~\ref{Fig:6_MoNbTaW_append_data}\textbf{e},\textbf{f} show the per‑atom strain fields at the end of tensile loading for the two cases. In the RSS case, deformation twinning initiates at the right crack tip, whereas brittle cleavage nucleates at the left tip. The twin boundary then interacts with the opposite (non‑crack) grain boundary, triggering a new crack. The twin band thickens and the GB migrates substantially to accommodate plastic deformation. These mechanisms are consistent with the sustained stress beyond 8\% strain in Fig.~\ref{Fig:6_MoNbTaW_append_data}\textbf{d}. In contrast, SRO markedly alters the deformation pathway: twinning is suppressed, only two dislocations are emitted from the crack in the presence of Nb segregation at the crack tip and the GB, and complete cleavage fracture is observed, as shown in Fig.~\ref{Fig:6_MoNbTaW_append_data}\textbf{f}. 

To validate the reliability of these simulations, we monitor the per‑atom extrapolation grade throughout the tensile deformation. As shown in Fig.~\ref{FigS-HEA-general}f,g, most of atomic environments remain within the interpolation domain of the training data, indicating that no significant extrapolation occurs. This confirms that \(\mathrm{ACE}_{\mathrm{UAMD,g}}\) reliably describes complex deformation involving GBs, dislocations, deformation twinning, and cracks.

\section*{DISCUSSION}
The rapid advancement of uMLIPs has transformed computational materials science, enabling high‐throughput simulations across diverse applications. Although many studies have demonstrated uMLIP accuracy in specific systems, validation remains challenging and circular: one needs DFT calculations to assess uMLIP predictions, yet the availability of reliable DFT data would obviate the need for uMLIPs. This dilemma highlights the urgent requirement for DFT-free UQ methods at every stage, from initial development through fine-tuning and model distillation. To address this, we propose an ensemble learning strategy that combines architecturally diverse uMLIP models with varying performance levels. By assigning weights to each ensemble member based on its test‐error statistics on the comprehensive OMat24 dataset, we construct a universal and sustainable uncertainty metric that can be applied to general inorganic material system. 

Critically, the proposed metric \(U\) not only provides a universal and reliable indicator of uMLIP prediction error but also enables a demonstrably sustainable approach to atomistic modelling. First, our workflow avoids the need to train or calibrate new uMLIP models, thereby eliminating the GPU‑years of computation and kilolitres of cooling water that modern AI training typically requires. Second, by reusing the more than twenty uMLIPs already available in the MatBench Discovery repository, we leverage their existing carbon footprint instead of creating redundant emissions. Third, the weighting scheme in Eq.~\ref{Eq:weighted_UQ} allows any future, higher‑accuracy uMLIP to be incorporated simply by evaluating its test errors on the OMat24 dataset, improving the uncertainty metric without additional training cycles. Finally, when combined with the UAMD protocol, \(U\) eliminates the vast majority of new DFT calculations: we demonstrate a 96\% reduction in DFT use for W potential development and complete avoidance of DFT in the expanded MoNbTaW dataset. Because the computational cost of running the eleven‑member uMLIP ensemble (Fig.\ref{Fig:2_UQ_on_OMat24}\textbf{}{a}) is negligible compared to DFT~\cite{shuang2025universal}, our method significantly reduces both the GPU/CPU compute time for training and the electricity consumption of energy‐intensive DFT calculations, offering a truly low‑carbon pathway for generating accurate interatomic potentials at scale.

Our UAMD framework unlocks the full potential of sMLIPs for atomistic modelling by addressing key limitations of existing model distillation approaches. Current model distillation from uMLIPs offers no mechanism to filter or correct errors inherited from the teacher~\cite{DPA2-distillation}. Fine‑tuning can mitigate this issue by retraining on new data~\cite{wang2025pfd,gardner2025distillation}, but it still demands substantial additional DFT calculations and GPU‑intensive training, and it carries the risk of catastrophic forgetting, which can degrade the original capability of uMLIPs~\cite{kim2025efficient}. In contrast, UAMD uses uncertainty estimates to selectively supplement the teacher’s labels with targeted DFT references, minimizing both numerical error and computational cost while preserving the extrapolation ability of uMLIPs. 

Our case studies on W and MoNbTaW alloys reveal several critical insights on model distillation. First, the accuracy of the student model (sMLIP) hinges on the fidelity of the uMLIP‐generated labels, while stochastic fitting errors during sMLIP training tend to cancel out against random data noise. Therefore, it is essential to choose the most accurate uMLIP as the surrogate for DFT and to replace any configurations with high uncertainty with true DFT calculations. Although uMLIPs are inherently slower than sMLIPs~\cite{shuang2025universal}, their computational cost can be mitigated by subsequent distillation into lightweight sMLIPs. Consequently, prioritizing uMLIP accuracy outweighs further improvements in computational efficiency. 
Second, UAMD can outperform direct training on raw DFT data (as illustrated in Fig.~\ref{Fig:4_W_system}\textbf{c}) because uMLIP predictions are inherently smoother owing to their advanced network architectures. Raw DFT labels contain nonphysical numerical noise~\cite{DFT-error-npj,DFT-error-JPCA}. The expressive power and regularization built into uMLIPs effectively filter out these fluctuations, yielding higher‑quality surrogate labels for sMLIP training. This denoising effect is particularly valuable for challenging configurations~\cite{Denoise}, such as large‑scale or highly distorted structures, where uMLIP predictions can even surpass the fidelity of DFT calculations. 

Additionally, our uncertainty metric \(U\) enables three complementary capabilities beyond model distillation, addressing both immediate practical needs and long‑term development of uMLIPs. First, by embedding uncertainty estimation into each prediction, outputs are accompanied by a uncertainty measure and configurations with \(U\) above a user‑defined threshold can be flagged for targeted DFT recalculation, ensuring reliability in critical simulations such as defect energetics or phase transformations. Second, \(U\) facilitates efficient fine‑tuning by identifying only the highest‑$U$ configurations for additional DFT labels. Third, systematic UQ‑driven dataset expansion leverages \(U\) to discover novel structures beyond existing dataset (for example, OMat24), guiding the gradual construction of ever more comprehensive training sets. Over successive cycles, this approach promises to converge on a truly universal potential that delivers near‑DFT fidelity across a broad spectrum of materials challenges.


In summary, this study develops an error-weighted UQ metric via the heterogeneous ensemble approach. Our validation across the diverse DFT datasets highlights its exceptional performance in estimating the prediction errors of uMLIPs without any DFT calculations and additional model training. We particularly show that the derived UAMD can unleash the powerfulness of machine learning in atomistic modeling by avoid the unaffordable computational cost of DFT, yet own the comparable accuracy. The potent applications of the new UQ method in the uMLIPs are discussed, further emphasizing the significance of our work.

\section*{METHODS}
\subsection*{Atomic cluster expansion potential development}
We employ pacemaker~\cite{drautz2019atomic} to develop ACE potentials for bcc W and MoNbTaW. For ACE, we adopt a highly nonlinear per‑atom energy in nonlinear form \(E_i=\varphi+\sqrt{\varphi}+\varphi^{1/8}+\varphi^{1/4}+\varphi^{3/8}+\varphi^{3/4}+\varphi^{7/8}+\varphi^{2}\), following Ref.~\cite{SiO-ACE-NC}. For the ACE basis, we use 72 functions (801 parameters) for W and 3{,}656 functions (30{,}868 parameters) for MoNbTaW. As the radial basis, we employ Bessel functions. During fitting, we set the force‑to‑energy weight ratio to \(\kappa=0.01\). We optimize the models using the BFGS algorithm for 2{,}000 steps. The cutoff distance is \(5~\text{\AA}\) for both W and MoNbTaW.

\subsection*{DFT calculations}
We use the Vienna \emph{ab initio} Simulation Package (VASP)~\cite{Kresse1996} to perform DFT calculations for W dimer and short‑range configurations. Exchange–correlation effects are described within the generalized‑gradient approximation using the Perdew–Burke–Ernzerhof (PBE) functional~\cite{Perdew1996}. Electron–ion interactions are treated with the projector‑augmented‑wave (PAW) method using the standard VASP PAW datasets. Electronic self‑consistency is converged to \(10^{-6}\,\mathrm{eV}\), and the plane‑wave cutoff is \(520\,\mathrm{eV}\). \(k\)-point meshes are generated with VASPKIT~\cite{VASPKIT} using Monkhorst–Pack grids, with a reciprocal‑space resolution of \(2\pi\times 0.03~\text{\AA}^{-1}\) applied consistently across the dataset.

\subsection*{Atomistic simulations}
We perform all atomistic simulations with LAMMPS~\cite{LAMMPS}. Atomic configurations are visualized and post‑processed with OVITO~\cite{stukowski2009visualization} (e.g., dislocation analysis). Together, these tools provide an integrated workflow for investigating the vacancy diffusion barriers, mechanical properties, and plastic‑deformation mechanisms of BCC W and MoNbTaW alloys.

\subsubsection*{Tensile simulation for bicrystals}
We perform bicrystal tensile simulations under fully periodic boundary conditions, with GB models constructed following Ref.~\cite{zheng2020grain}. The \(\Sigma 3\) tilt GB \(\left(109.47^\circ\,[110]\,(1\overline{1}2)\right)\) model measures \(66 \times 203 \times 108~\text{\AA}^3\) and contains approximately 90{,}000 atoms. The \(\Sigma 3\) twist GB \(\left(60.00^\circ\,[111]\,(111)\right)\) model measures \(85 \times 210 \times 74~\text{\AA}^3\) and contains about 82{,}000 atoms. In both cases, the separation between GBs exceeds \(10~\mathrm{nm}\).
Before loading, each system is energy‑minimized and equilibrated at \(300~\mathrm{K}\) for \(20~\mathrm{ps}\). Uniaxial tension is applied normal to the GB plane at an engineering strain rate of \(5 \times 10^8~\mathrm{s}^{-1}\), while the temperature is maintained at \(300~\mathrm{K}\) using a Nosé–Hoover thermostat. For the tilt GB model, deformation proceeds to \(50\%\) engineering strain. For the twist GB model, the simulation is terminated upon brittle fracture.

We perform similar tensile simulations for MoNbTaW bicrystals with and without an initial crack. For the crack‑free models shown in Fig.~\ref{Fig:5_MoNbTaW}f, the initial bicrystal measures \(132 \times 203 \times 108~\text{\AA}^3\) and contains 18{,}000 atoms. The bicrystal incorporates a  \(\Sigma 3\ 109.47^\text{o}[110](1\overline{12})\) tilt GB. The strain rate and loading manner is the same the simulations in pure W. For the cracked case in Fig.~\ref{Fig:6_MoNbTaW_append_data}e, the simulation cell measures \(280 \times 207 \times 46~\text{\AA}^3\) and contains 154{,}750 atoms, retaining the same \(\Sigma 3\) tilt GB and loading protocol. The initial crack measures \(60 \times 9.5 \times 46~\text{\AA}^3\). 

\subsubsection*{MC/MD simulations}
The MC/MD simulations are conducted to generate chemical short range order (SRO) at 300 K by LAMMPS \cite{LAMMPS}. The samples are initially relaxed and equilibrated at 300 K and zero pressure under the isothermal-isobaric (NPT) ensemble through MD. After that, MC steps consisting of attempted atom swaps are conducted, hybrid with the MD. In each MC step, a swap of one random atom with another random atom of a different type is conducted based on the Metropolis algorithm in the canonical ensemble. 100 MC swaps are conducted at every 1000 MD steps with a time step of 0.001 ps during the simulation. $1\times10^6$ steps are conducted in MC/MD simulations.



\section*{Data availability}
The data that support the findings of this study are available upon reasonable request. The parameter files of the potential and training dataset will be published upon the publication of this paper.

\section*{Code availability}
The data that support the findings of this study are available at the GitHub repository: https://github.com/Kai-Liu-MSE/UQ-uMLIP.

\section*{Acknowledgments}
This work was sponsored by Nederlandse Organisatie voor WetenschappelijkOnderzoek (The Netherlands Organization for Scientific Research, NWO) domain Science for the use of supercomputer facilities. The authors also acknowledge the use of DelftBlue supercomputer, provided by Delft High Performance Computing Center (https://www.tudelft.nl/dhpc).

\section*{Author Contributions}
K.L. \& F.S.: Conceptualization; Data Curation; Formal Analysis; Investigation; Methodology; Project Administration; Software; Validation; Visualization; Writing—Original Draft; Writing—Review \& Editing;
Z.W.: Data Curation; Software; Writing;
W.G. \& P.D. \& M.S.: Writing—Review \& Editing, Supervision

\section*{Conflict of Interest}
The authors declare no conflict of interest.

\newpage
\section*{Figures}

\newpage

\newpage

\newpage

\newpage

\newpage

\clearpage 

%
\bibliography{ref} 
\bibliographystyle{sciencemag}

%
%
%
%
%
%



\newpage


\renewcommand{\thefigure}{S\arabic{figure}}
\renewcommand{\thetable}{S\arabic{table}}
\renewcommand{\theequation}{S\arabic{equation}}
\renewcommand{\thepage}{S\arabic{page}}
\setcounter{figure}{0}
\setcounter{table}{0}
\setcounter{equation}{0}
\setcounter{page}{1} 


\begin{center}
\section*{Supplementary Materials for\\ \scititle}
\end{center}
\subsubsection*{This PDF file includes:}

Supplementary Note 1-2

\noindent Supplementary Table S1 and S2

\noindent Supplementary Figure S1 to S6

\newpage

\newpage
\subsection*{Supplementary Note 1. Details of extensive DFT datasets for $U$ validation}
The datasets used in this study exhibit broad diversity in elemental coverage, configurational variety, and application scenarios.
\begin{enumerate}
    \item \textbf{Elemental diversity}: the datasets span chemistries from elemental metals (e.g., TM23) and binaries (e.g., FeC, LiH) to complex multicomponent alloys such as MoNbTaVW, as well as hybrid systems including MOF, ICOF, and perovskites. Several datasets (e.g., MOF, Perovskite) extend beyond metallic systems to include ionic, covalent, and organic–inorganic frameworks.
    \item \textbf{Configurational diversity}: unlike the OMat24 dataset, which contains only periodic bulk structures and explicitly excludes point defects, surfaces, non‑stoichiometry, and lower‑dimensional motifs~\cite{barroso2024open}, the datasets used here sample a far broader structural space. They include liquids (TM23, HfO$_2$), amorphous and disordered phases (C, Ga$_2$O$_3$), surfaces and catalytic models (surface catalyst, PtH), clusters and small molecules (e.g., dimers and trimers in Fe), and high‑energy configurations generated by space‑group enumeration (e.g., Mg-spg). This expanded configurational breadth, coupled with the uncertainty metric \(U\), enables effective identification of high‑uncertainty structures well beyond the scope of OMat24.
    \item \textbf{Application scenarios}: the datasets support a wide range of scientific problems, including solid‑state deformation in metals and alloys, phase transformations (e.g., Perovskite), diffusion and thermal transport (e.g., FeH, HfO$_2$, ICOF), catalytic processes (e.g., PtH, surface catalyst), and battery electrode materials (e.g., LiPS, LiPO, Na$_3$V$_2$(PO$_4$)$_3$).
\end{enumerate}
Together, these attributes provide high coverage across chemical, structural, and functional spaces, yielding a dataset suite with broad, near‑universal applicability. As a result, the proposed uncertainty $U$ evaluated here are positioned to generalize across diverse bonding types, length scales, and operating conditions, rather than being confined to narrow materials classes or single application domains.
 
On the other hand, because the datasets originate from heterogeneous sources, the underlying DFT labels are not uniform. Protocol differences include \(k\)-point mesh density, plane‑wave cutoff energy, electronic convergence thresholds, smearing schemes, the choice of pseudopotentials/PAW datasets and exchange–correlation functionals, the inclusion or omission of Hubbard \(U\) or dispersion corrections, and whether spin polarization (collinear or non‑collinear) is treated. For example, the TM23 and M16 datasets are generated without spin polarization, which is problematic for magnetic systems. In practice, for ferromagnetic elements such as Fe and Co included in TM23, the resulting DFT reference errors can exceed the uncertainty metric \(U\); in such cases, cross‑source inconsistency dominates the error budget, inflating apparent test errors and obscuring the relationship between \(U\) and true model uncertainty. Moreover, several subsets are produced with different electronic‑structure codes. For example, the Perovskite and surface catalyst datasets are computed with Quantum ESPRESSO (QE), which introduces additional differences in pseudopotential formats, smearing/convergence behavior, and energy referencing relative to VASP‑labeled data. Despite heterogeneity in composition, structure, and DFT protocols, the proposed uncertainty metric \(U\) exhibits a stable and significant correlation with the true errors, enabling consistent risk ranking across datasets and delivering a principled basis for UAMD and other applications.

\begin{figure}[htbp]
    \centering
    \includegraphics[width=.45\linewidth]{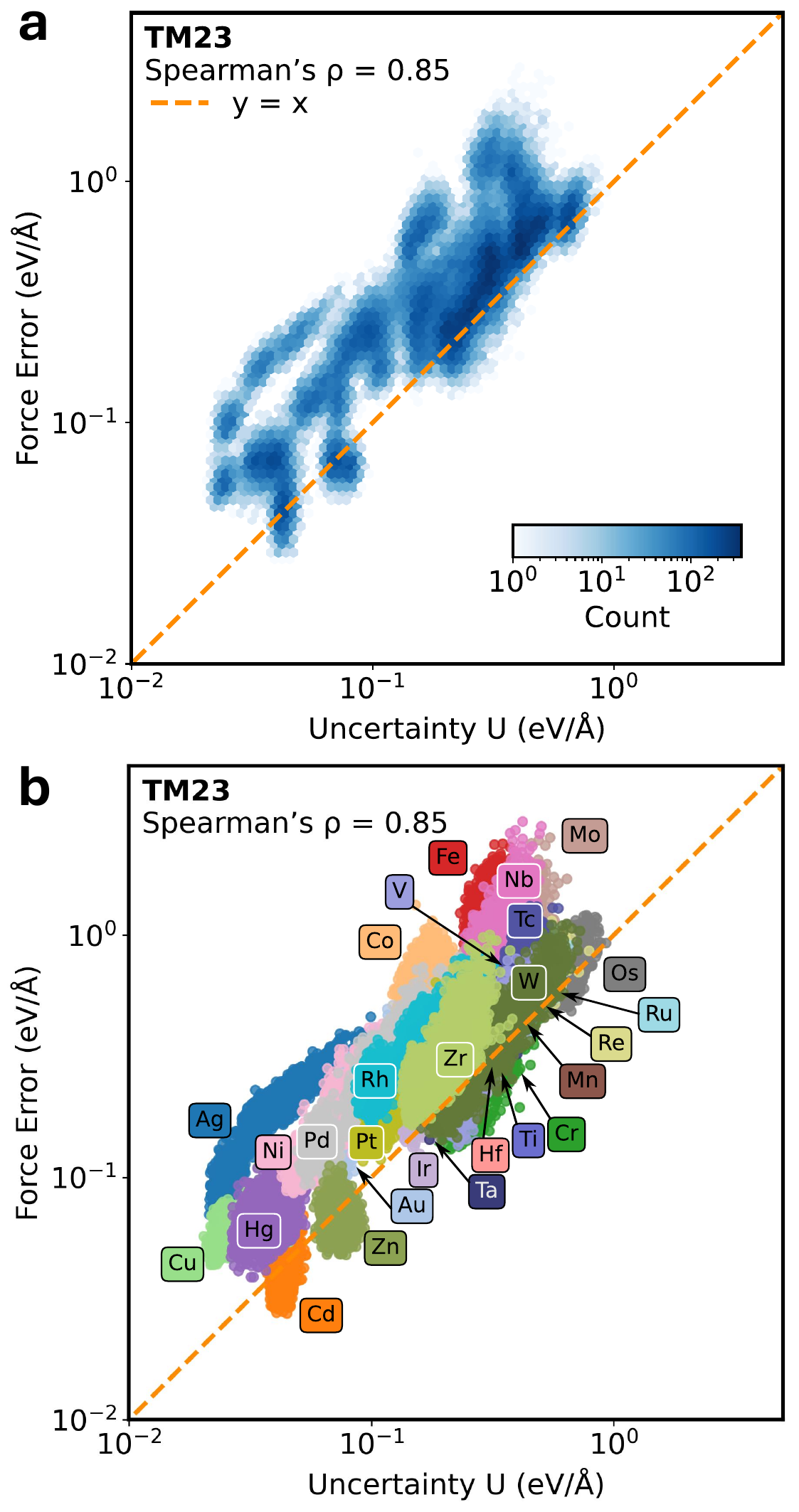}
    \caption{\textbf{Parity plots of uncertainty \(U\) versus force prediction error for the TM23 dataset~\cite{owen2024complexity}.} 
The TM23 dataset includes only pure metals. 
\textbf{a} Hexbin plot with color indicating point density. 
\textbf{b} Scatter plot with points colored by element type. 
For some elements, data points overlap; their approximate locations are indicated by arrows.}
    \label{Fig:S1}
\end{figure}

\begin{figure}[htbp]
    \centering
    \includegraphics[width=0.5\linewidth]{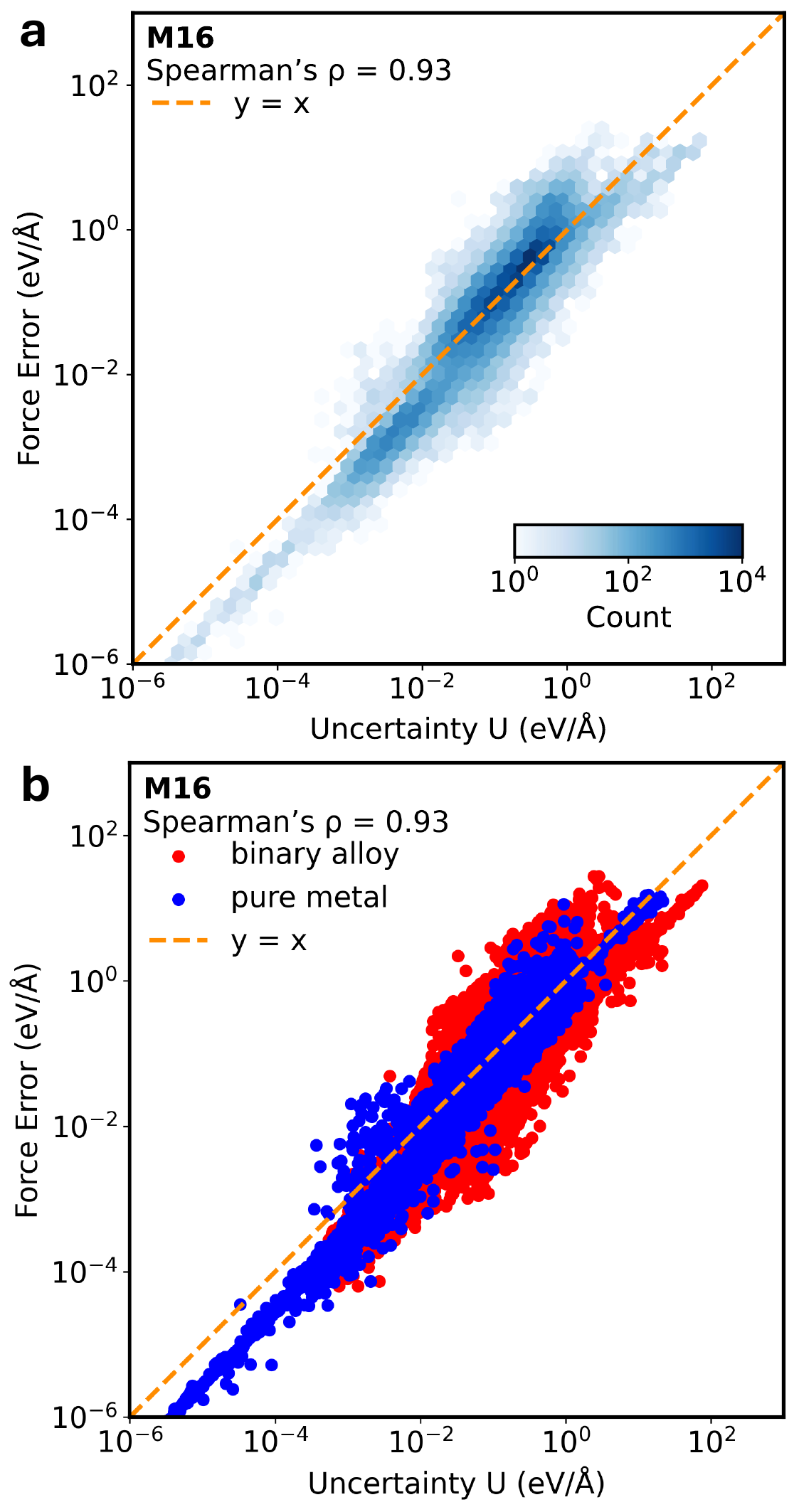}
    \caption{\textbf{Parity plots of uncertainty \(U\) versus force prediction error for the M16 dataset.} 
The M16 dataset includes both pure metals and binary alloys. 
\textbf{a} Hexbin plot with color indicating point density. 
\textbf{b} Each configuration type is indicated by color: red for binary alloys and blue for pure metals.}
    \label{Fig:S2}
\end{figure}

The TM23 and M16 datasets cover a broad range of elements in the metallic and alloy systems and exhibit substantial configurational diversity. They contain 81,000 and 105,464 configurations, respectively, including a large number of liquid-phase structures. Fig.~\ref{Fig:S1} and \ref{Fig:S2} show the correlation between the uncertainty metric \(U\) and the force prediction error for TM23 and M16, with Spearman’s \(\rho\) reaching 0.85 and 0.93, respectively. In Fig.~\ref{Fig:S1}\textbf{b}, where the individual elements are labeled, it is evident that the correlation between \(U\) and error varies significantly with element type, consistent with the discussion in TM23 regarding the effect of electronic localization on the difficulty of potential fitting. Fig.~\ref{Fig:S2}\textbf{b} presents the \(U\)–error relationship for pure metals and binary alloys, showing that binary alloys generally exhibit higher \(U\) values and larger errors than pure metals. It should be noted that the DFT calculations for both datasets were performed without spin polarization, which is inconsistent with the typical training datasets used for uMLIPs. This discrepancy likely explains the systematic underestimation of error by \(U\) observed in Fig.~\ref{Fig:S1}\textbf{b} for magnetic elements such as Fe, Co, and Nb.


\newpage
\subsection*{Supplementary Note 2. Dataset construction for developing general-purpose potential for MoNbTaW alloys}
We begin with the defect genome for bcc W from Ref.~\cite{shuang2025modeling} and generate chemically diverse variants by (i) rescaling the lattice parameter to that of the target random alloy, (ii) reassigning species to satisfy the target composition, and (iii) sampling with the Maximum‑Volume algorithm implemented in MLIP‑2~\cite{MLIP-2}. The W defect‑genome set contains 384 distinct topologies (e.g., dislocations, surfaces, and crack tips). To enhance transferability across chemistries, we construct a composition grid with 10\% increments over the Mo-Nb-Ta-W space, yielding 287 target compositions that span the full quaternary space. Instantiating each composition for every topology produces \(287 \times 384 = 110{,}208\) candidate structures, which is prohibitively large for direct DFT labeling.

We therefore down‑select the pool using the command select-add in MLIP‑2's routine, which implements a maximum‑volume (MaxVol) criterion in the descriptor space to identify a diverse and representative subset. Unless otherwise noted, we use the default select-add settings and operate on Moment Tensor Potential (MTP) computed for the entire pool. This procedure yields 4{,}696 configurations that preserve coverage of the major defect classes (dislocation cores, GB/crack‑tip environments, and free surfaces) and the breadth of compositions on the 10\% grid, while substantially reducing the labeling cost.

To improve robustness near the edges of composition space, we additionally include reference datasets for the elemental bcc metals (Mo, Nb, Ta, and W), the equiatomic quaternary \(\mathrm{Mo}_{25}\mathrm{Nb}_{25}\mathrm{Ta}_{25}\mathrm{W}_{25}\), and the binary random alloy \(\mathrm{Mo}_{50}\mathrm{Nb}_{50}\). The final training pool comprises 7{,}000 configurations, combining a compositionally stratified, defect‑rich subset from the genome expansion with baseline elemental and benchmark‑alloy data. This design yields an ACE model that is transferable across structures and compositions while avoiding the full DFT labeling cost.

\newpage

\subsection*{Supplementary Tables}

\renewcommand{\arraystretch}{0.9}
\setlength{\tabcolsep}{4pt}

\begin{table}[ht]
  \centering
  \begin{tabularx}{\textwidth}{@{}
      >{\hsize=1.5\hsize\raggedright\arraybackslash}X  
      >{\hsize=0.5\hsize\centering\arraybackslash}X    
      >{\centering\arraybackslash}X    
      >{\centering\arraybackslash}X    
      >{\centering\arraybackslash}X    
      >{\centering\arraybackslash}X    
    @{}}
    \toprule
    uMLIP                   & Ref.                         & \makecell{Num.\ of\\parameters} & \makecell{Force RMSE\\(eV/\AA)} & \makecell{Energy RMSE\\(eV/atom)} & \makecell{$w_{k}$ in $U$,\\Eq.~\ref{Eq:weighted_UQ}} \\
    \midrule
    eSEN‑30M‑omat           & \cite{fu2025learning}& 30M& 0.237& 0.223&0.12182                 \\ 
    eqV2‑31M‑omat           & \cite{barroso2024open}& 31M& 0.243& 0.191&0.11877                 \\
    eqV2‑86M‑omat           & \cite{barroso2024open}& 86M& 0.244& 0.178&0.11866                 \\
    SevenNet‑omat           & \cite{kim2024data}& 26M& 0.254& 0.241&0.11389                 \\
    SevenNet‑mf‑ompa‑omat & \cite{kim2024data}& 26M& 0.264& 0.245&0.10932                 \\
    SevenNet‑mf‑ompa‑mpa    & \cite{kim2024data}& 26M& 0.278& 0.259&0.10412                 \\
    MACE‑omat               & \cite{batatia2023foundation}&  9M& 0.350& 0.246&0.08250\\
    Orb‑v3‑c‑20‑omat        & \cite{rhodes2025orb}& 25M& 0.488& 0.251&0.05925                 \\
    Orb‑v3‑c‑inf‑omat       & \cite{rhodes2025orb}& 25M& 0.498& 0.295&0.05807            \\
    Orb‑v3‑d‑20‑omat        & \cite{rhodes2025orb}& 25M& 0.506& 0.250&0.05706                 \\
    Orb‑v3‑d‑inf‑omat       & \cite{rhodes2025orb}& 25M& 0.511& 0.251&0.05654                 \\
    eqV2‑31M‑OAM            & \cite{barroso2024open}& 31M& 0.779& 0.297&-          \\
    eqV2‑86M‑OAM            & \cite{barroso2024open}& 86M& 0.817& 0.297&-               \\
    eSEN‑30M‑OAM            & \cite{fu2025learning}& 30M& 1.077& 0.281&-               \\
    GRACE‑2L‑omat           & \cite{bochkarev2024graph}& 13M& 1.384& 1.367&-               \\
    GRACE‑2L‑OAM            & \cite{bochkarev2024graph}& 13M& 1.591& 0.514&-               \\
    DPA3‑V2                 & \cite{zhang2025graph}&  7M& 2.462& 0.311&-               \\
    DPA3‑V1                 & \cite{zhang2025graph}&  8M& 6.221& NaN&-                 \\
    \bottomrule
  \end{tabularx}
  \caption{The table summarizes the uMLIPs used in this work, including their parameter counts and performance on the OMat24 test set. For force RMSE, force values are clipped at 1000~eV/\AA. The DPA3‑V1 model produces erroneous energy predictions for some configurations, resulting in NaN values.}
  \label{tab:uMLIP}
\end{table}

\newpage

\renewcommand{\arraystretch}{0.8}
\setlength{\tabcolsep}{4pt}

\begin{table}[htbp]
  \centering
  \begin{tabularx}{\textwidth}{@{} 
      >{\raggedright\arraybackslash}X  
      >{\centering\arraybackslash}X    
      >{\raggedright\arraybackslash}X  
      >{\centering\arraybackslash}X    
      >{\centering\arraybackslash}X    
    @{}}
    \toprule
    Name              & Ref.                                & Elements                                           & Num.\ of Conf. & Group \& Note               \\
    \midrule
    CrCoNi            & \cite{sheriff2024quantifying}      & Cr, Co, Ni                                         & 1257           & Alloy (PBE)               \\
    Fe                & \cite{jana2023searching}           & Fe                                                 & 10627          & Metal (PBE)               \\
    MoNbTaTi          & \cite{MoNbTaTi}                                 & Mo, Nb, Ta, Ti                                     & 18870          & Alloy (PBE)               \\
    MoNbTaVW          & \cite{MoNbTaVW}                                 & Mo, Nb, Ta, V, W                                   & 9848           & Alloy (PBE)               \\
    VCoNi             & \cite{VCoNi}                                 & V, Co, Ni                                          & 6311           & Alloy (PBE)               \\
    WTaCrV            & \cite{WTaCrV}                                 & W, Ta, Cr, V                                       & 13796          & Alloy (PBE)               \\
    Mg spg            & \cite{poul2023systematic}          & Mg                                                 & 17210          & Metal (PBE)               \\
    TM23              & \cite{owen2024complexity}           & 27 d‑block metals (Ti–Hg)                          & 81000          & Metal (PBE)               \\
    M16               & \cite{song2024general}              & Ag, Al, Au, Cr, Cu, Mg, Mo, Ni, Pb, Pd, Pt, Ta, Ti, V, W, Zr & 105464         & Alloys (PBE)               \\
    FeH               & \cite{meng2021general}              & Fe, H                                              & 11298          & IC (PBE)  \\
    MoNbTaWH          & \cite{shuang2025decoding}           & Mo, Nb, Ta, V, H                                   & 9999           & IC (PBE)  \\
    LiH               & \cite{LiH}                                 & Li, H                                              & 603            & IC (PBE)                   \\
    FeC               & \cite{meng2024highly}               & Fe, C                                              & 20578          & IC (PBE)                   \\
    HfO$_2$           & \cite{sivaraman2020machine,zhang2023vibrational} & Hf, O                             & 3876           & IC (PBE)                   \\
    Ga$_2$O$_3$       & \cite{zhao2023complex}              & Ga, O, C                                        & 1628           & IC (PBE)                   \\
    GaN               & \cite{sun2025heat}                  & Ga, N                                              & 1110           & IC (PBE)                   \\
    PtH               & \cite{vandermause2022active}        & Pt, H                                              & 250            & Other (PBE)                   \\
    C                 & \cite{wang2025density}              & C                                                  & 7188           & Other (PBE)                   \\
    ICOF              & \cite{li2025decoding}               & H, B, C, O, Li, Na                                 & 1392           & Other (PBE)                   \\
    MOF               & \cite{yue2024toward}                & C, H, O, N, Zn                                                  & 19705          & Other (PBE+D3)                   \\
    Perovskite        & \cite{shi2023investigation}         & Cs, Pb, Cl, Br, I                                  & 4200           & Other (QE)                   \\
    LiPS              & \cite{park2021accurate}             & Li, P, S                                           & 3000           & Other (PBE)                   \\
    Li$_4$P$_2$O$_7$  & \cite{park2021accurate}                                 & Li, P, O                                                  & 3000           & Other (PBE)                   \\
    Surface catalyst  & \cite{schaaf2023accurate}           & H, C, O, In, Pt                                    & 2273           & Other (QE)                   \\
    Na$_3$V$_2$(PO$_4$)$_3$ 
                      & \cite{de2023nanosecond}             & O, Na, P, V                                        & 23456          & Other (PBE+U)                   \\
    \bottomrule
  \end{tabularx}
  \caption{Summary of DFT datasets used to validate the proposed uncertainty metric $U$. "PBE" refers to standard VASP calculations, "+D3" includes dispersion corrections, "+U" indicates Hubbard corrections, and "QE" denotes calculations performed with Quantum ESPRESSO, and "IC" refers to inorganic compounds.}
  \label{tab:DFT-valid}
\end{table}

\FloatBarrier 

\newpage

\subsection*{Supplementary Figures}

\newpage

\newpage
\begin{figure}[htbp]
    \centering
    \includegraphics[width=1\linewidth]{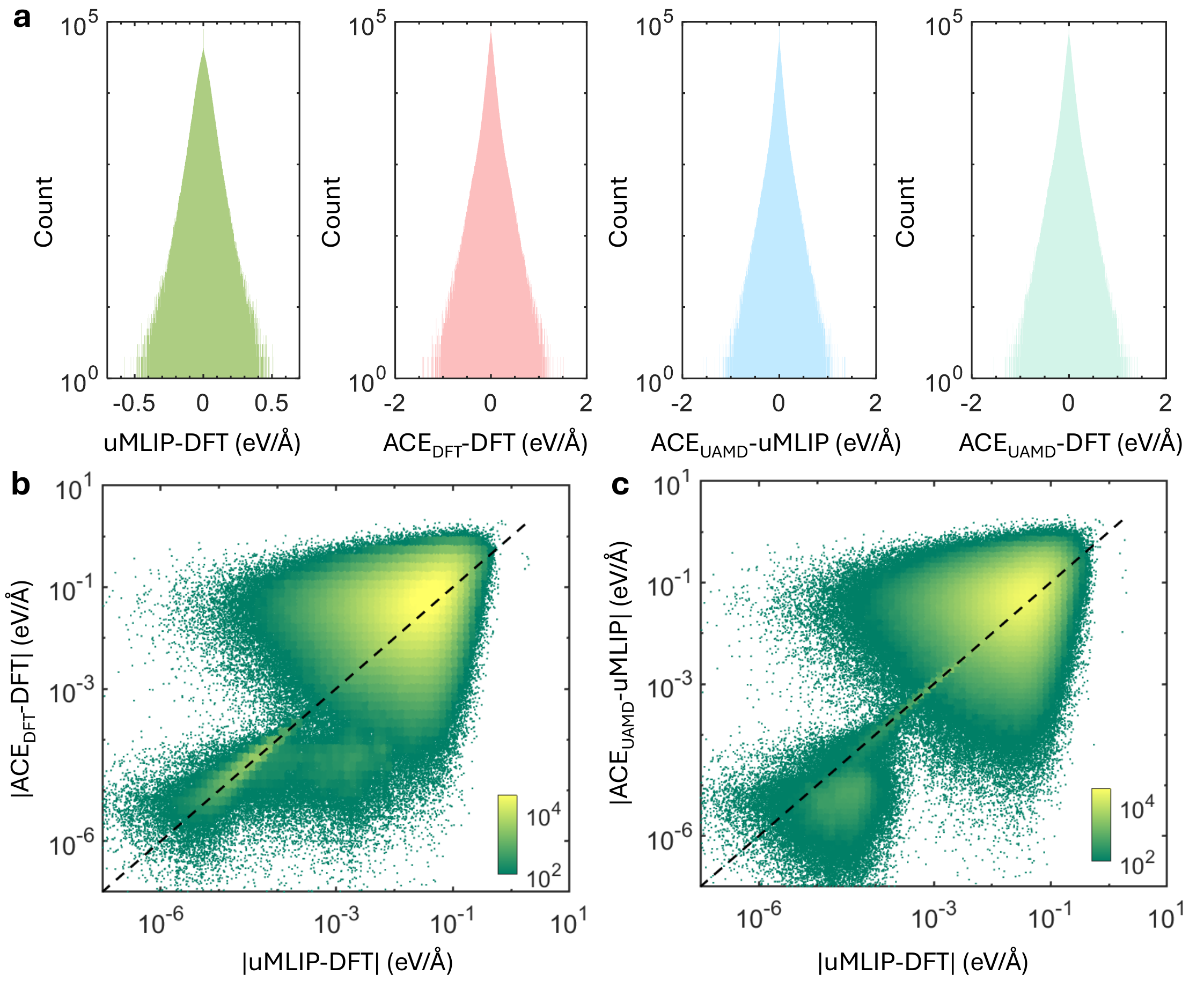}
    \caption{\textbf{Error analysis in UAMD of MoNbTaW alloys.} 
    \textbf{a} Histogram plot of various force errors. 
    \textbf{b} Prediction errors of DFT-derived ACE model versus uMLIP prediction errors. 
    \textbf{c} Prediction errors of UAMD-derived ACE model versus uMLIP prediction errors.}
\label{Fig:error_correlation}
    \label{FigS-UAMD-HEA-error}
\end{figure}

\newpage
\begin{figure}[htbp]
    \centering
    \includegraphics[width=1\linewidth]{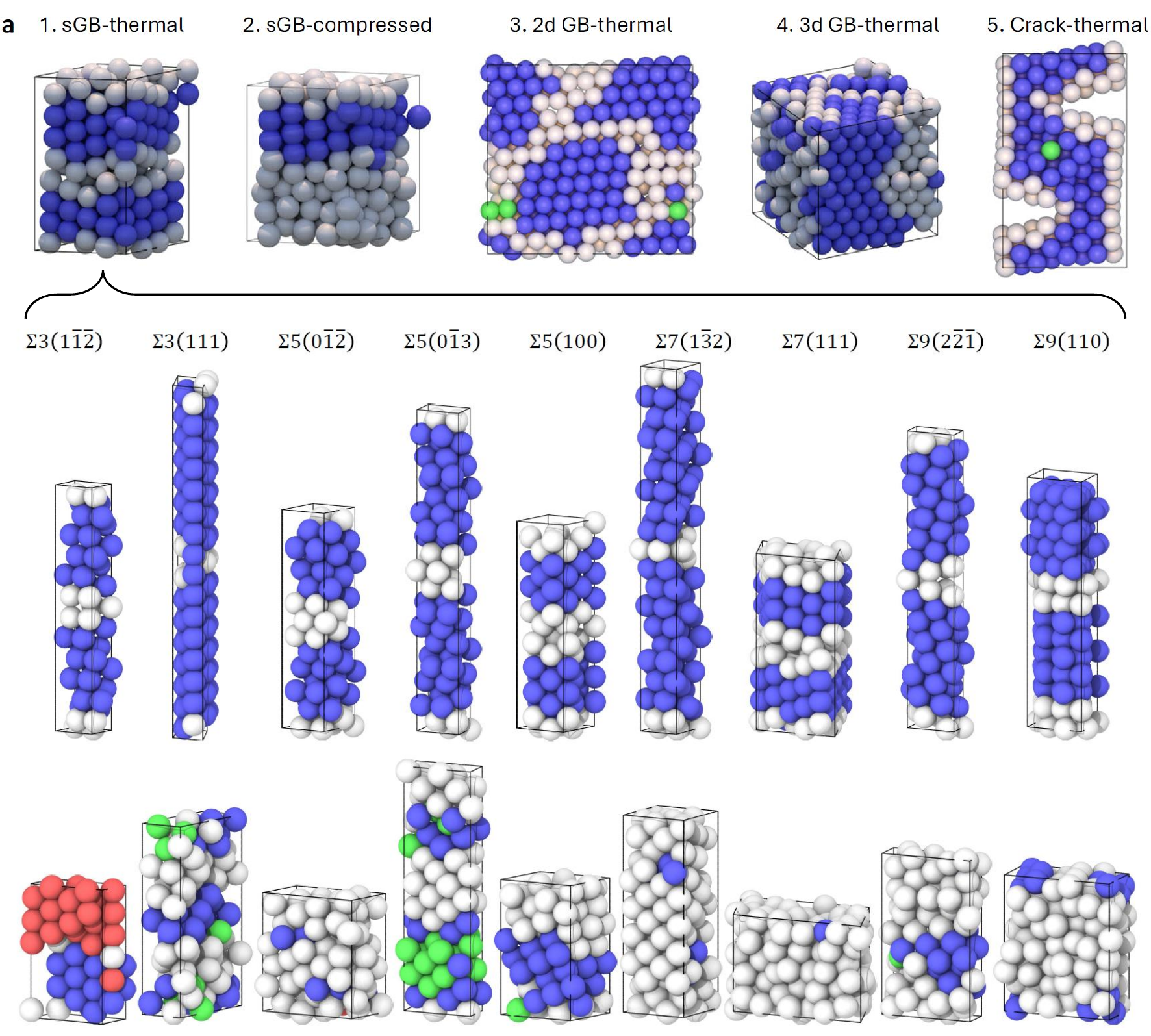}
    \caption{\textbf{Validation datasets sourced from Ref.\cite{shuang2025modeling} for W.}}
    \label{FigS-npj-test}
\end{figure}

\newpage
\begin{figure}[htbp]
    \centering
    \includegraphics[width=1\linewidth]{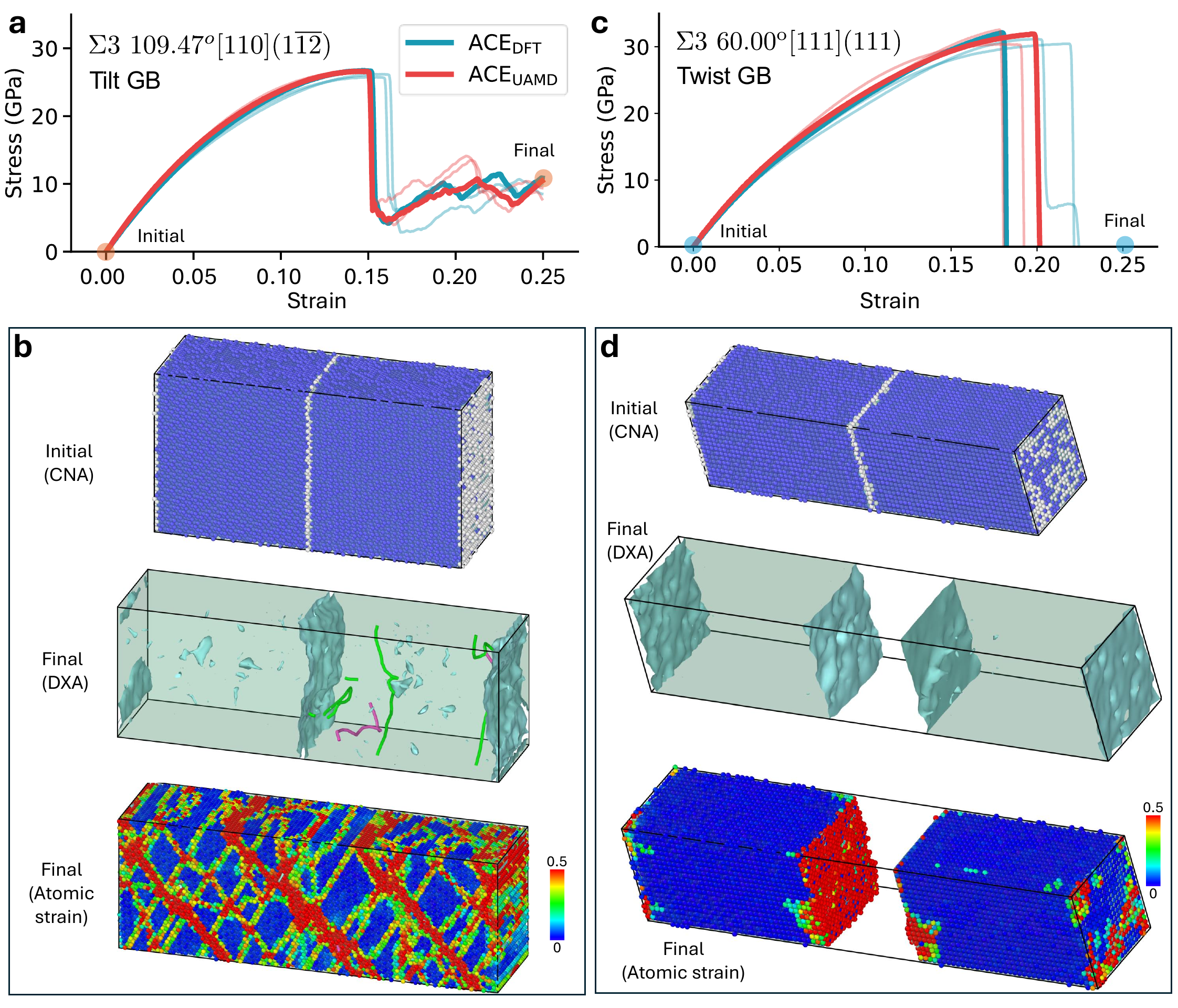}
    \caption{\textbf{Details of the MD simulations for W bicrystal tensile tests using the distilled $\text{ACE}_{\text{UAMD}}$  potential.} 
    \textbf{a,c} Stress-strain curves for the tilt and twist GBs. 
    \textbf{b,d} Initial atomic configurations (visualized by Common Neighbor Analysis, CNA), final dislocation structures (Dislocation Extraction Analysis, DXA), and atomic shear strain distributions for the two bicrystals.}
    \label{FigS-W-deformation}
\end{figure}

\newpage
\begin{figure}[htbp]
    \centering
    \includegraphics[width=0.8\linewidth]{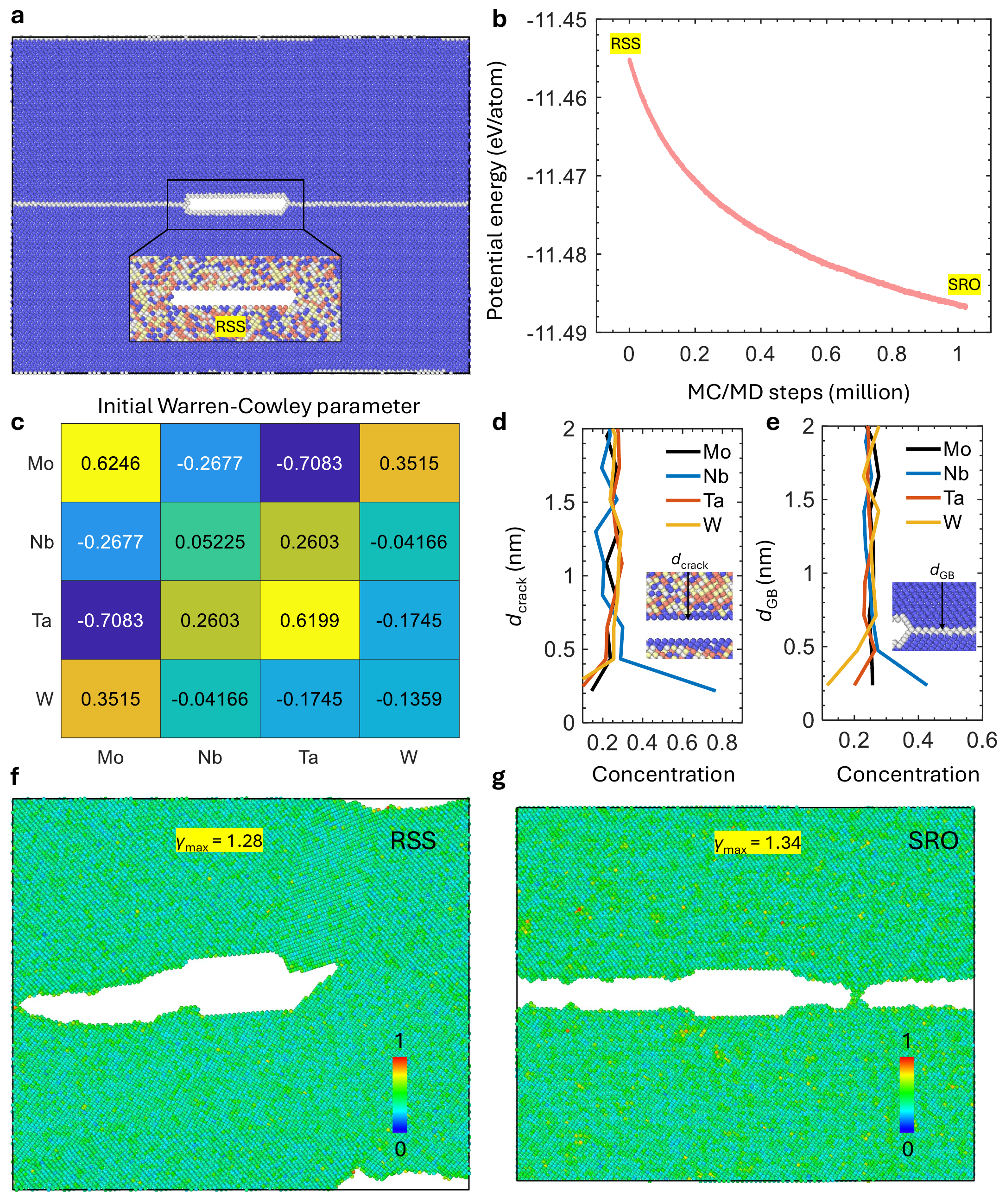}
    \caption{\textbf{Details of the MD simulations for Mo\textsubscript{25}Nb\textsubscript{25}Ta\textsubscript{25}W\textsubscript{25} bicrystal fracture using the general-purpose potential ($\text{ACE}_{\text{UAMD,g}}$).}
    \textbf{a} Bicrystal model used for simulations, with atoms color-coded by Common Neighbor Analysis (CNA). The inset shows the random solid solution of Mo\textsubscript{25}Nb\textsubscript{25}Ta\textsubscript{25}W\textsubscript{25}. 
    \textbf{b} Energy evolution during MC/MD simulation. 
    \textbf{c} Initial Warren-Cowley (WC) parameters before tensile loading. 
    \textbf{d,e} Elemental distribution near the crack surface and grain boundary (GB). 
    \textbf{f,g} Per-atom extrapolation grade ($\gamma$) at the end of the tensile simulation, with the maximum grade ($\gamma_\text{max}$) during deformation labeled for each case.  Values of $\gamma \leq 1$ indicate interpolation, while $1 < \gamma < 2$ corresponds to safe extrapolation.}

    \label{FigS-HEA-general}
\end{figure}

\end{document}